%% file: f.tex
\documentclass[12pt]{article}
\usepackage{epsf}
\usepackage{cite}
\usepackage{subeqnarray}

\input{ambjorn}

\input{macros.tex}

\begin{document}
\setlength{\topmargin}{0pt}
\setlength{\oddsidemargin}{5mm}
\setlength{\headheight}{0pt}
\setlength{\headsep}{0pt}
\setlength{\topskip}{9mm}

\input{f_tit}

\input{f_all}

\newpage 
\input{f_tab}
\clearpage
\newpage 
\input{f_fig}
\clearpage

\end{document}

%% file: ambjorn.tex
\newcommand{\rf}[1]{(\ref{#1})}
\newcommand{\bea}{\begin{eqnarray}}
\newcommand{\eea}{\end{eqnarray}}
\newcommand{\e}{{\rm e}}
\newcommand{\g}{\gamma}

\renewcommand{\b}{\beta}
\renewcommand{\a}{\alpha}

\newcommand{\del}{\delta}
\newcommand{\D}{\Delta}
\renewcommand{\L}{\Lambda}

\newcommand{\ep}{\varepsilon}

\newcommand{\sg}{\sigma}

\newcommand{\vph}{\varphi}
\newcommand{\oh}{\frac{1}{2}}
\newcommand{\oq}{\frac{1}{4}}

\newcommand{\ra}{\right\rangle}
\newcommand{\la}{\left\langle}

\newcommand{\cD}{{\cal D}}

\def\void{}
\def\labelmark{}

\newenvironment{formula}[1]{\def\labelname{#1}
\ifx\void\labelname\def\junk{\begin{displaymath}}
\else\def\junk{\begin{equation}\label{\labelname}}\fi\junk}%
{\ifx\void\labelname\def\junk{\end{displaymath}}
\else\def\junk{\end{equation}}\fi\junk\labelmark\def\labelname{}}

{\ifx\void\labelname\def\junk{\end{array}\end{displaymath}}
\else\def\junk{\end{array}\right.\end{equation}}
\fi\junk\labelmark\def\labelname{}\def\junk{}
}

\newcommand{\beq}{\begin{formula}}
\newcommand{\eeq}{\end{formula}}
\newcommand{\beqv}{\begin{formula}{}}
\newcommand{\bes}{\begin{subeqnarray}}
\newcommand{\ees}{\end{subeqnarray}}

%% file: macros.tex
\newcommand{\vev}[1]{\langle {#1} \rangle}

%% file: f_tit.tex
\begin{flushright}
\hfill    NBI-HE-96-69\\
\hfill    December 96\\
\end{flushright}

\begin{center}
\vspace{24pt}
{\large \bf Quantum geometry of 2d gravity coupled to unitary matter}

\vspace{24pt}

{\sl J. Ambj\o rn } and {\sl K. N. Anagnostopoulos}

\vspace{6pt}

 The Niels Bohr Institute\\
Blegdamsvej 17, DK-2100 Copenhagen \O , Denmark\\

\vspace{12pt}

\end{center}
\vspace{24pt}

\vfill

\begin{center}
{\bf Abstract} 

\end{center}

\vspace{12pt}

\noindent
We show that there exists a divergent correlation length in $2d$
quantum gravity for the matter fields close to the critical point
provided one uses the invariant geodesic distance as the measure of
distance. The corresponding reparameterization invariant two-point
functions satisfy all scaling relations known from the ordinary theory
of critical phenomena and the KPZ exponents are determined by the
power-like fall off of these two-point functions. The only difference
compared to flat space is the appearance of a dynamically generated
fractal dimension $d_h$ in the scaling relations. We analyze
numerically the fractal properties of space--time for Ising and
three--states Potts model coupled to $2d$ dimensional quantum gravity
using finite size scaling as well as small distance scaling of
invariant correlation functions. Our data are consistent with $d_h=4$,
but we cannot rule out completely the conjecture $d_H =
-2\alpha_1/\alpha_{-1}$, where $\alpha_{-n}$ is the gravitational
dressing exponent of a spin-less primary field of conformal weight
$(n+1,n+1)$. We compute the moments $\vev{L^n}$ and the loop--length
distribution function and show that the fractal properties associated
with these observables are identical, with good accuracy, to the pure
gravity case.

\vfill

\newpage

%% file: f_all.tex
\section{Introduction}

\subsection{General framework}

Two-dimensional gravity has been intensely studied the last six 
years, both as a toy model for four-dimensional gravity and because 
of its importance in string theory. Let the partition function 
of two-dimensional quantum gravity coupled to a conformal field theory
be defined by 
\beq{*in1}
Z(\L) = \int \cD [g] \cD_g \phi \; \e^{-\L \int d^2\xi \sqrt{g} - S_M(\phi,g)},
\eeq
where $\L$ denotes the cosmological constant, 
the integration is over equivalence classes of metrics $[g]$,
and $S_M(\phi,g)$ is the matter Lagrangian. Furthermore, if $\vph$ is 
a primary conformal field of scaling dimension $\D_0$ in the theory 
defined by $S_M(\phi)$ in flat space and if  $\la\,( \cdot)\, \ra_M$ denotes
the functional average of an observable $( \cdot)$ calculated with 
the action $S_M(\phi)$, we have 
\beq{*in2}
\la \vph(\xi_1)\vph(\xi_2) \ra_M \sim \frac{1}{|\xi_1 - \xi_2|^{2\D_0}}.
\eeq
If $V$ denotes the volume of flat space-time Eq.\ \rf{*in2} implies
the finite size scaling 
\beq{*in3}
\int_V d^2\xi\int_V d^2\xi' \la \vph(\xi)\vph(\xi') \ra_M \sim V^{2-\D_0}
\eeq
for $V$ sufficiently large. In the seminal papers of KPZ and DDK 
\cite{kpz,ddk} it was shown
how to extend this result to the situation where the conformal field theory 
was coupled to two-dimensional quantum gravity. Let $Z(V)$ be the 
partition function \rf{*in1}, only restricted to universes of 
space-time volume $V$, viz.\
\beq{*in6}
Z(V) = \int \cD [g] \cD_g \phi \; \e^{ - S_M(\phi,g)}\, \del(
\int d^2\xi \sqrt{g} -V),
\eeq
and let $\la\, (\cdot)\,\ra_{M+G}$ denote the functional average 
of an observable $(\cdot)$ in the ensemble defined by $Z(V)$.
With this notation the finite size scaling relation of KPZ and DDK reads: 
\beq{*in4}
\la \int d^2\xi \sqrt{g}\int d^2\xi' \sqrt{g} \; \vph(\xi)\vph(\xi')
\ra_{M+G} \sim V^{2-\D},
\eeq
where the exponent $\D$ is related to $\D_0$ via 
\beq{*in5}
\D = \frac{\sqrt{1-c+24\D_0}-\sqrt{1-c} }{\sqrt{25-c}-\sqrt{1-c}}.
\eeq
In \rf{*in5} $c$ denotes the central charge of the conformal field
theory defined in flat two-dimensional space-time by $S_M(\phi)$.
Furthermore it was shown that the partition function for fixed volume
behaves as
\beq{*in7}
Z(V)\sim V^{\g -3},~~~~~\g = \frac{c-1-\sqrt{(c-1)(c-25)}}{12}.
\eeq

We want to emphasize that all these results refer to the partition
function or to correlators which are integrated over all space-time.
In flat space-time the scaling properties of the two-point functions
\rf{*in2} are considered the underlying reason for the \emph{ finite size 
scaling relations} such \rf{*in3}. Clearly Eqs.\ \rf{*in4} and \rf{*in7}
are corresponding finite size scaling relations in the conformal field 
theories coupled to quantum gravity and it is natural to expect that 
the scaling properties are dictated by two-point functions depending on 
the geodesic distance. Nevertheless the analogy of \rf{*in2} 
in quantum gravity has only recently been analyzed 
\cite{aw,syracuse,ajw}, despite the fact that 
the two-point functions depending on the \emph{ geodesic distance} probe
the metric properties of space-time in a much more direct way. In fact, 
as we shall see, the two-point functions are perfect probes of the 
fractal structure of quantum space-time, and they will highlight the fact 
that even if we start out with an underlying two-dimensional 
manifold, there will be metric properties of two-dimensional quantum 
space-time which cannot be viewed as two-dimensional.

A basic property of the continuum two-point function \rf{*in2} (for 
infinite volume) is its invariance under translations and rotations. 
If the volume $V$ is sufficiently large we can write:
\beq{*in8}
G^{(0)}_\vph (R;V) \equiv \frac{1}{V}       
\int_V d^2\xi\int_V d^2\xi' \;\frac{\la \vph(\xi)\vph(\xi') \ra_M\, 
\del ( |\xi-\xi'| - R )}{n_0(R;V)}\sim \frac{1}{R^{2\D_0}}, 
\eeq
where $n_0(R;V)$ denotes the volume of a spherical shell of radius $R$.
By definition we have from \rf{*in3} that 
\beq{*in8a}
\int dR\; n_0(R;V)\, G^{(0)}_\vph (R;V) \sim V^{1-\D_0}.
\eeq
While it is difficult to generalize Eq.\ \rf{*in2} to two-dimensional 
quantum gravity since our physical observables have to be reparameterization
invariant, Eq.\ \rf{*in8} has a simple translation to quantum gravity:
\beq{*in9}
G_\vph (R;V) \equiv
\la \frac{1}{V} \int d^2\xi \sqrt{g} \int d^2 \xi'\sqrt{g} \;
\frac{\vph(\xi) \vph(\xi') \del (D_g(\xi,\xi')-R)}{n_g(\xi;R;V)} \ra_{M+G}.
\eeq
In this equation the average is defined via the partition function 
\rf{*in6} for metrics of volume $V$. $D_g(\xi,\xi')$ denotes the geodesic 
distance between $\xi$ and $\xi'$ with respect to the metric $g$, and
$n_g(\xi;R;V)$ is the volume of a spherical shell\footnote{Notice that
such a spherical shell is not necessarily connected.} of geodesic
radius $R$ with center at $\xi$:
\beq{*in11}
n_g(\xi;R;V) = \int d^2\xi' \sqrt{g} \; \del (D_g(\xi,\xi')-R).
\eeq
We can define the average volume of a spherical shell, 
or the volume-volume correlator,  $n(R;V)$ by 
\beq{*in12}
n(R;V) = \la \frac{1}{V} \int d^2\xi \sqrt{g} \;n_g(\xi;R;V) \ra_{M+G},
\eeq
and from dimensional analysis we expect that the analogue of Eq.\ \rf{*in8a}
will be 
\beq{*in13}
\int_0^\infty dR \;n(R;V)\, G_\vph (R;V) \sim V^{1-\D}\, ,
\eeq
where $\D$ is the KPZ exponent in \rf{*in4}. It is sometimes convenient 
to use the ``unnormalized'' correlator 
\beq{*in13a}
n_\vph (R;V) = \la \frac{1}{V}
\int d^2\xi \sqrt{g} \int d^2 \xi' \sqrt{g}\;
\vph(\xi) \vph(\xi')\, \del (D_g(\xi,\xi')-R) \ra_{M+G}
\eeq 
rather than $G_\vph(R;V)$. In this case we have by definition
\beq{*in13b}
\int_0^\infty dR \; n_\vph (R;V) \sim V^{1-\D},
\eeq
and for $\vph =1$ we obtain $n_1(R;V) = n(R;V)$, justifying the 
name volume-volume correlator for the average value of $n_g(\xi;R;V)$.
From the definition of $n(R;V)$ and in accordance with $\D = 0$ for 
$\vph =1$, we have 
\beq{*in14}
\int_0^\infty dR \;n_1 (R;V) \sim V,
\eeq  
and if we \emph{ assume} that 
\beq{*in15}
n_1(R;V) \sim R^{d_h-1}~~~\textrm{for}~~~R \ll V^{1/d_h},
\eeq
where we denote $d_h$ the \emph{ intrinsic Hausdorff dimension}
or the \emph{fractal dimension}  
of quantum space-time, it is natural from Eq.\ \rf{*in13} to expect
\beq{*in16}
G_\vph (R;V) \sim \frac{1}{R^{d_h \D}}~~~\textrm{for}~~~R \ll V^{1/d_h},
\eeq
\beq{*in16a}
n_\vph (R;V) \sim \frac{R^{d_h-1}}{R^{d_h\D}}
~~~\textrm{for}~~~R \ll V^{1/d_h}.
\eeq
\emph{ Eq.\ \rf{*in16} becomes the quantum gravity version of Eq.\
\rf{*in8a}.}

The main goal of this article is to provide evidence for Eqs.\ 
\rf{*in15}-\rf{*in16a} and to determine $d_h$ as precisely as possible.
However, at this point let us point to a subtlety in the definition of
$d_h$.  It is possible provide an additional definition of the fractal
dimension of space-time, starting from \rf{*in1} rather than
\rf{*in6}. In this situation we do not keep the space-time volume
fixed. Instead we define a ``global'' fractal dimension $d_H$ by
\beq{*global1}
\la V \ra_R \equiv \la \int d^2 \xi \sqrt{g} \ra_R \sim R^{d_H}
\eeq
for $R \leq \L^{-1/d_H}$. The average in \rf{*global1} is over 
all universes where two marked points are separated a geodesic distance $R$,
i.e.\ calculated from the partition function
\beq{*global2}
Z_R(\L) = 
\int \cD [g] \cD_g \phi \; \e^{-\L \int d^2\xi \sqrt{g} - S_M(\phi,g)}
\int d^2\xi \sqrt{g}\int d^2\xi'\sqrt{g} \;\del (D_g(\xi,\xi')-R).
\eeq 
In pictorial terms the calculation of \rf{*global1} corresponds to 
the calculation of the average length of a random walk which has traveled
a distance $R$. A priori there is no reason 
for $d_h =d_H$ and it is indeed possible\footnote{The so--called 
multicritical branched polymer models are  examples where $d_h = 2$ while 
$d_H = m$, $m=2,3, \ldots.$}  to find statistical models where $d_h \neq d_H$. 
However, as we shall see, for two-dimensional gravity coupled to conformal 
matter it seems that there is only one fractal dimension, valid at all 
distances. As a consequence, the constant coefficients in Eqs.\
\rf{*in15}{--}\rf{*in16a} are independent of the volume $V$. 

\subsection{Present status}
 
Before discussing in more details the model and the methods used, let us 
briefly review what is known and what is conjectured so far.

Using the so-called transfer matrix formulation \cite{transfer}
it has been possible to calculate $Z_R(\L)$ in the case 
of pure gravity. It is remarkably simple \cite{aw}:
\beq{*simple}
Z_R(\L)= \L^{3/4} \frac{\cosh (\sqrt[4]{\L} R)}{\sinh^3 (\sqrt[4]{\L}R )}.
\eeq
From the definition of $Z_R(\L)$ it follows that it is related to 
$n_1(R;V)$ by
\beq{*simple1}
Z_R(\L) = \int_0^\infty dV \e^{-\L V}\, VZ(V) \, n_1(R;V),
\eeq
and it is easy to prove that
\beq{*in17}
n_1(R;V) = R^3 f_1(x),~~~~x=\frac{R}{V^{\oq}},
\eeq
where $f_1(x)$ can be expressed in terms of generalized 
hyper-geometric functions and
\beq{*in18}
f_1(0) > 0 ,~~~~~~f_1(x) \sim \e^{-x^{4/3}}~~\textrm{for} ~~ x \gg 1.
\eeq
From \rf{*in15} we conclude that $d_h=4$ in the case of 
pure gravity, while \rf{*simple} implies that $d_H=4$.
Hence  $x$ is a dimensionless scaling variable. 
It is useful to write Eq.\ \rf{*in17} as 
\beq{*in19}
n_1(R;V) = V^{1-1/d_h} F_1(x), 
\eeq
\beq{*in20}
F_1(x) \sim x^{d_h-1}~~\textrm{for}~~ x \ll 1.
\eeq
These equations appear as typical finite size scaling relations, 
ideally suited  for numerical simulations. In the case of pure gravity 
one can calculate additional correlators of the so-called gravitational 
descendents \cite{ope}. 

The final observable which, in the case of pure gravity, can be calculated 
from first principles is the so-called loop distribution function. 
If $P$ denotes a point on the two-dimensional manifold of volume $V$,
we consider the set of points whose geodesic distance from $P$ is 
equal $R$. Generically, this set consists of several closed 
loops of various length $L_i$. We take the average over positions $P$
as well as over the ensemble of all two-dimensional spherical 
manifolds of volume $V$ (with weight $e^{-S_M(\phi,g)}$ if matter fields 
are included). In this way we get a distribution $\rho_V(R,L)$ such that 
$\rho_V(R,L)dL$ measures the average number of loops with length 
between $L$ and $L + dL$ per manifold with volume $V$. 
We can now define the moments of $L$:
\beq{*in32}
\la L^n (R)\ra_V = \int_0^\infty dL \; L^n \, \rho_V(R,L).
\eeq
Notice that 
\beq{*in33}
n_1(R,V) = \la L (R) \ra_V.
\eeq
In the case of pure gravity it is possible, using again the 
transfer matrix, to calculate $\rho_V(R,L)$ in the 
limit where $V \to \infty$ \cite{transfer}:
\beq{*in34}
\rho_\infty (R,L) = 
\frac{1}{R^2} \;\hat{\rho} (L/R^2),
\eeq
where 
\beq{*in35}
\hat{\rho}(y) = \textrm{const.} \times \Bigl( y^{-5/2} + \oh y^{-1/2} 
+ \frac{14}{3} y^{1/2} \Bigr) \; \e^{-y}.
\eeq
One observes the singularity of $\hat{\rho}$ for small $y$. This 
is why a cut-off $\epsilon$ in $L$ is needed. In fact, $\la L^0 \ra$ and 
$\la L \ra$ are singular. If we cut off the integral in \rf{*in32}
at $\epsilon$ (the lattice length unit of $L$) we obtain:
\bea
\la L (R)\ra_{V = \infty} &=&
\frac{\textrm{const.}}{\sqrt{\epsilon}} 
\, R^3 + O(1) , \label{*in36}\\
\la L^n(R) \ra_{V = \infty} &=& c_n \, R^n,~~~~~n >1. \label{*in37}.
\eea
If the volume $V$ is finite \rf{*in37} is replaced by 
\beq{*in37a}
\la L^n(R) \ra_V = V^{2n/d_h} F_n(x),
\eeq 
where $d_h=4$ and where 
\beq{*in37b}
F_n(x) \sim x^{2n}~~~{\rm for}~~x \ll 1.
\eeq
Since we know that $L$ has dimension of length and $d_h =4$, 
i.e.\ $R$ has dimension $[L^{1/2}]$, relation \rf{*in37} is 
dimensionally correct, but it is clear that we need a 
dimensionful parameter in the relation $\la L (R)\ra \sim R^{d_h-1}$.
It can only be provided by the cut-off $\epsilon$.

It is natural to conjecture \cite{syracuse,ajw} that finite size
scaling relations like \rf{*in19} and \rf{*in20} are valid also for
unitary conformal field theories coupled to quantum gravity, except
that $d_h$ could be a function of the central charge $c$ of the
conformal field theory. Furthermore, it has been conjectured that the
same $d_h(c)$ and the same scaling variable $x=R/V^{1/d_h(c)}$ govern
the finite size scaling relations of all the correlators $G_\vph(R;V)$
of primary conformal fields $\vph$ of the given conformal field
theory:
\beq{*in21}
G_\vph(R;V) = \frac{1}{R^{d_h\D}} \, \tilde{g}_\vph (x) = V^{-\D} g_\vph(x),
\eeq
where the scaling function $g_\vph(x)$ then behaves as
\beq{*in22}
g_\vph(x) \sim x^{-d_h\D}~~~\textrm{for}~~~x\ll 1,
\eeq
and 
\beq{*in22ab}
g_\vph(x) \to 0~~~\textrm{for}~x \to \infty.
\eeq
It is sometimes convenient to use the ``unnormalized'' correlation functions 
instead:
\beq{*in21a}
n_\vph(R;V) = \frac{R^{d_h-1}}{R^{d_h\D}} \,\tilde{F}_\vph (x) 
= V^{1-\D-1/d_h} F_\vph(x),
\eeq
where 
\beq{*in22a}
F_\vph(x) \sim x^{d_h-1 -d_h\D}~~~\textrm{for}~~x\ll 1.
\eeq 

A priori it is not clear what to expect for $\rho_V(R,L)$ if 
conformal matter is coupled to quantum gravity. Rather surprisingly,
it seems that the finite size scaling relation \rf{*in37b} is still
valid with $d_h=4$ replaced by $d_h(c)$, 
and that $\hat{\rho} (y)$ is still a function only of $L/R^2$
when $V \to \infty$. In case $d_h(c)$ is different from 4 this implies
that $L$ has an anomalous scaling dimension relative to $V$. 
This has been shown convincingly to be true for the $c=-2$ model
in \cite{dgi}. In this paper we show that the numerical data is
consistent with Eq.~\rf{*in37a} and Eq.~\rf{*in37b} by extracting
consistent values for $d_h$ from \rf{*in37a} and by showing that
\rf{*in37b} holds with very good accuracy. We should warn the reader
though, that since $d_h\approx 4$ in our case, the scaling
\rf{*in37a}{--}\rf{*in37b} is an assumption based on the $c=-2$
result.

It should be emphasized that the assumption of a single new parameter
$d_h(c)$ which determines uniquely all scaling in the given conformal
field theory coupled to quantum gravity is a strong assumption which
so far has avoided a rigorous mathematical proof. There has very few  
attempts to discuss analytically the appearance of a divergent
correlation length when, say, a spin system coupled to gravity becomes
critical \cite{ha}.  However, the scaling has
already been tested numerically for the Ising model and the
three-states Potts model coupled to gravity \cite{syracuse,ajw,aamt}.
Recall that the unitary conformal field theories with central charge
$c$ between 0 and 1 are the so-called minimal $(q,p)$ rational conformal
field theories with $(q,p)=(m,m+1)$, $m=2,3,\ldots$ and the
corresponding central charge is $c=1-6/m(m+1)$.  Pure gravity
corresponds to $m=2$, i.e.\ $c=0$, the Ising model corresponds to
$m=2$, i.e.\ $c=1/2$ and the three-states Potts model corresponds to
$m=5$ and $c=4/5$.  Using $n_1(R;V)$ and $n_\vph(R;V)$
it was found that $d_h \approx 4$ independent of these three values of
$c$, while $d_h\D$ was found to be consistent with the value predicted
by KPZ, provided $d_h$ was attributed a value close to 4 \cite{aamt}.
In this way the numerical simulations have so far provided some
support for the scaling hypothesis outlined above, but with the rather
surprising result that $d_h \approx 4$ for $0 \leq c < 1$. This is in
contradiction with the two theoretical predictions for $d_h$ which
exist so far:
\beq{*in23}
d_h^{(i)} = 2 \times \frac{\sqrt{25-c}+\sqrt{49-c}}{\sqrt{25-c}+
\sqrt{1-c}}
\eeq
and
\beq{*in24}
d_h^{(ii)} = \frac{24}{\sqrt{1-c}\;(\sqrt{1-c}+\sqrt{25-c})}~~~~
\Bigl( = -\frac{2}{\g} \Bigr).
\eeq
The first formula \cite{watabiki,noboru} is derived under certain assumptions 
about the scaling of the diffusion equation in Liouville theory
and predicts a rather slow change in $d_h$ as a function of $c$. The 
second formula, derived for $(q,p)$ models coupled to quantum gravity
by means of string field theory \cite{kawai}, 
is based on the assumption that the 
``time'' in string field theory can be identified with geodesic distance.
It is true by construction for $c=0$, but might not be true for $c \neq 0$.
Note in particular that $d_h^{(ii)}(c) \to \infty$ for $c \to 1$ and 
$d^{(ii)}_h(c) \to 0$ for $c \to -\infty$. In the table below 
we have summarized the predictions made so far, including numerical 
results too.

\vspace{12pt}

\begin{center}
\begin{tabular}{| c | c | c | c |}
\hline
c     & $d^{(i)}_h$ & $d^{(ii)}_h$ & numerical \\ \hline
$-\infty$ & 2      &   0        &  - \\
-2    & 3.562      &   2        & $3.58 \pm 0.04$ \\
0     & 4          &   4        & $\approx 4$     \\
1/2   & 4.21       &   6        & $\approx 4$     \\
4/5   & 4.42       &   10       & $\approx 4$     \\
1     & 4.52       & $\infty$   & $\approx 4$    \\ \hline
\end{tabular}
\end{center}

\vspace{12pt}
\noindent
Let us emphasize the numerical results for $c=-2$ \cite{dgi} (see also
\cite{kketal} for earlier measurements). For $c=-2$ one can avoid the 
use of Monte Carlo simulations since there exists a recursive
algorithm which directly generates independent configurations. The
statistics are for this reason very good and the numerical results
clearly favor the prediction $d_h^{(i)}$. In addition, the measurement
of $\rho_V(R,L)$ for $c=-2$ supports \rf{*in37a} and \rf{*in37b} with
$d_h(c)= d_h^{(i)}(c)$ and $L/R^2$ as dimensionless scaling variable.

\subsection{Outline}
 It is the aim of the present article to discuss the current available
precision in the determination of $d_h(c)$ by the use of numerical
simulations and to test the scaling hypothesis
\rf{*in37a}-\rf{*in22a}.  We perform simulations for the Ising model
coupled to gravity and the three-states Potts model coupled to gravity
as well as simulations for pure gravity (as a check of the
accuracy). In particular we measure the volume-volume correlator
$n_1(R;V)$, the moments $\vev{L^n(r)}_V$ ($n=2,3,4$) as well as the
spin-spin correlators $G_\vph(R;V)$ and $n_\vph(R;V)$ and apply finite
size and small distance scaling in order to determine $d_h(c)$.  The
existence of such scaling provides evidence that there is a unique
fractal dimension $d_h(c)$ which extends over all distance, a highly
non-trivial fact for the critical spin systems coupled to quantum
gravity. Furthermore we measure the so-called loop length distribution
function $\rho_V(R,L)$ which provides additional insight in the
fractal structure of quantum gravity. By comparing the different
measurements (which are more or less consistent) we get a handle on
the \emph{systematic} finite size effects which affect the
determination of $d_h(c)$.

In the next section we introduce our discretized model of quantum
gravity, the so-called dynamical triangulation model of quantum
gravity, and discuss how to extract the observables defined above
within the framework of dynamical triangulations. In particular we
discuss the so-called shift $a$ which appears when the observables
measured on a finite lattice is mapped to the continuum observables
defined above. In section 3 we describe the numerical method and
present our results. Finally we discuss our results in section 4.

\section{The model}

We define the regularized theory of two-dimensional quantum gravity via 
the formalism known as dynamical triangulations \cite{dtr}. Some of the 
exact results for pure gravity mentioned in the introduction were 
derived within this formalism which in certain respect is more 
powerful than a continuum formalism, even for analytical calculations.

In this formalism surfaces  
are constructed from equilateral triangles glued to together
to form triangulations with spherical topology\footnote{We will 
consider only surfaces with spherical topology in the following.}. 
We allow for the generation of certain degenerate triangulations.
They are described most easily by using the fact that the 
graphs dual to triangulations are $\phi^3$-graphs. Regular triangulations
correspond to connected $\phi^3$-graphs without tadpole and 
self energy sub-graphs. The degenerate triangulations include the
connected $\phi^3$ graphs with tadpoles and self energies.
The use of the 
full set of connected $\phi^3$-graphs of spherical topology is known to improve
scaling compared to the class of $\phi^3$-graphs which
corresponds to regular triangulations \cite{atw}. 
The coupling of spin systems to gravity is done by assigning
a spin to each  vertex in the triangulation, either Ising of three-states
Potts spin, depending on the model. At the critical temperatures 
for the spin systems on dynamical triangulations 
such models  describe  conformal field theories 
of central charge $c=1/2$ and $c=4/5$ coupled to gravity, respectively.
In this way  the discretized partition function corresponding to 
the formal continuum partition function \rf{*in6} is 
\beq{*in25}
Z(N) = \sum_{T_N} \sum_{\{ \sg_i\}} \e^{- S_{T_N}(\{\sg_i\})},
\eeq
where the first summation is over all (spherical) triangulations
$T_N$ consisting of $N$ triangles and where the second summation is 
over all spin configurations on $T_N$. 
$S_T(\{\sg_i\})$ denotes the action for the Ising or the
three-states Potts model defined on the triangulation $T$ in the 
way mentioned above, viz.\
\beq{*in25a}
S_T (\{\sg_i\}) = \b  \sum_{(k,l)} \sg_k \sg_l ,
\eeq 
where $\b$ is the temperature, the summation is over all pairs 
of neighboring vertices $(k,l)$ and $q$ refers to the $q$-states Potts
model ($q=2$ for the Ising model). 
For the $q$-states Potts model the spin variables can take $q$ values 
and $\sg_k \sg_l$ is a symbolic notation for 
\beq{*in25b}
\sg_k \sg_l \equiv 
\frac{q}{q-1} \del_{\sg_k,\sg_l} - \frac{1}{q-1}.
\eeq
The continuum volume $V$ is related the number of triangles $N$ by 
$$
V \propto  N \ep^2
$$
where $\ep$ is the length of the individual links in the triangulations.
As usual in lattice theories we work in units of $\ep$, i.e.\ $\ep =1$.
Phased entirely in the framework of statistical mechanics we study 
the annealed average of $q$-states Potts models  
on the class of all triangulations with spherical
topology. 

Geodesic distances $r$ on the triangulations are defined either as the 
shortest link distance between two vertices or as the shortest 
path through neighboring triangles. While these two distances 
can vary a lot for two specifically chosen vertices in a given 
triangulation, they are proportional when the average is performed 
over the ensemble of triangulations. We will report here the results 
obtained by the use of link distance, since it is known that the short distance
behavior of correlators suffer from less discretization effects if 
we use the link distance than if we use the triangle distance\footnote{
Intuitively it is to be expected that the link distance will 
behave more ``continuum like'' for short link distances compared to 
short distance triangle distances. At short distance the 
triangle distance is quite ``rigid'': each triangle has 3
neighboring triangles (except for some degenerate triangulations), 
while vertices can have a variable number of neighboring vertices.}
\cite{syracuse}.

The lattice analogue of the correlator functions 
$n_1(R;V)$, $n_\vph(R;V)$ and $g_\vph(R;V)$ can now be defined. We will
use the same notation for the function names, but $r$ and $N$ instead of 
the continuum variables $R$ and $V$. No confusion should be possible. 
Let $i$ and $j$ denote lattice vertices and $D_{ij}$ the (lattice) geodesic 
distance between $i$ and $j$ and $r$ an integer (the lattice 
analogy of $R$, measured in units of $\ep$). Then, 
\bea
n_1(r;N)& =& \frac{1}{N} \la \sum_i \sum_j \del_{D_{ij},r} \ra,
\label{*in26}\\
n_\vph (r;N) &=& \frac{1}{N} \la \sum_i \sum_j \sg_i \sg_j\, 
\del_{D_{ij},r} \ra,
\label{*in27}\\
G_\vph (r;N) &=&  \frac{1}{N} \la \sum_i \sum_j 
\frac{\sg_i\sg_j \, \del_{D_{ij},r}}{\sum_k \del_{D_{ik},r}} \ra,
\label{*in28}
\eea 
where $\sg_i\sg_j$ is defined as in \rf{*in25b}. For a given vertex $i$
we always perform the summation $\sum_j$ in 
Eqs.\ \rf{*in26}--\rf{*in28}. However, from a numerical point 
of view it is not convenient to perform the summation over $i$ for 
a given triangulation since the sums over $j$ for different 
choices of $i$ will be highly dependent. On the other hand it is not 
convenient either from the  point of view of numerical efficiency
to choose just a single vertex $i$ for each triangulation. We found it 
convenient to generate independent configurations (by Monte Carlo 
simulations) and for each such configuration to choose randomly 
a suitable number $N_0$ of vertices $i$ and replace 
$$
\frac{1}{N} \sum_{i=1}^N \to \frac{1}{N_0} \sum_{i_\a , \a=1}^{N_0}.
$$
We have checked that the numerical results are idependent of $N_0$.
In particular, one can choose the extreme value $N_0=1$, although it is 
not convenient from the point of view of efficiency, as mentioned above.

We now expect the following short distance behavior:
\beq{*in29}
n_1(r;N) \sim r^{d_h-1},~~~~n_\vph (r;N) \sim r^{d_h-1-d_h\D},~~~~
G_\vph (r;N) \sim r^{-d_h \D},
\eeq
as long as $r \ll N^{1/d_h}$, where the constant coefficients in the
above relations are independent of $N$. The scaling hypothesis can be
formulated as
\bea 
n_1(r;N)  &\sim&  N^{1-1/d_h} F_1(x),  \label{*in30}\\
n_\vph(r;N) &\sim&  N^{1-\D-1/d_h}F_\vph(x), \label{*in30a}\\
G_\vph(r;N) &\sim& N^{-\D} g_\vph (x), \label{*in30b}
\eea
where
\beq{*in31}
x =\frac{r}{N^{1/d_h}},
\eeq
and the functions $F(x)$, $F_\vph(x)$ and $g_\vph(x)$ should 
approach their continuum counter parts in the limit $N \to \infty$.
It is known that this approach can be slow \cite{syracuse,ajw}.
It can be improved by applying a typical finite size
scaling argument \cite{ajw}: we can only expect 
strict proportionality 
\beq{*inx1}
r \ep^{2/d_h} \propto R~~~\textrm{and}~~~~N \ep^2 \propto V
\eeq
for $N \to \infty$. For finite $N$ one expects
\beq{iii1a}
\frac{R}{V^{1/d_h}} = \textrm{const.}\times \frac{r}{N^{1/d_h}} +
O \Bigl(\frac{1}{N^{1/d_h}} \Bigr),
\eeq
since $1/N^{1/d_h}$ measures the linear extension of the system,
i.e.\ the simplest finite size correction to $x$ can be parameterized as
\beq{iii1}
x \equiv \frac{r+a}{N^{1/d_h}}\, ,
\eeq
where $a$ is called the ``shift''. In principle this shift might
depend on the observable we consider. This shift was first introduced
in \cite{ajw} in the case where $r$ is the triangle distance.  It was
shown that in the case of pure gravity a remarkable agreement between
the discrete and continuum volume-volume correlators can be achieved
for $a\sim 5.5$, even for quite small lattices. These authors
introduce a second shift variable $N^{1/d_h}\rightarrow N^{1/d_h}+b$
which in this study we will set equal to $0$. In the case of link
distance, as our results will show, $a$ is much smaller, in agreement
with the remark above concerning finite size effects for link and
triangle distances, and it is possible to obtain very good results
from finite size scaling without introducing it
\cite{syracuse,aamt,aamt2}. Its role, however, is quite important
when trying to determine $d_h$. One cannot ignore it since, as it was
shown in \cite{ajw}, it is a necessary finite size correction in the
case where one uses triangles to define geodesic distance and area in
the discretized version of $n_1(R;V)$. Its introduction improves the
fits in the case of link distance as well, but it will add an extra
free parameter, since we have no analytical results in the case of
matter coupled to gravity. This will make the determination of $d_h$
more difficult. Moreover, $a$ improves dramatically the small distance
scaling of the correlators, making it possible to perform direct fits
near the origin in order to extract the scaling exponents, which would
otherwise be impossible. 

Finally, the continuum definitions for the loop length distribution, 
Eqs.\ \rf{*in34}-\rf{*in37}, are valid at the discretized level 
with the replacement 
\beq{*in38}
R \to \textrm{const.} \times \, r,~~~~~
L \to \textrm{const.} \times\, l,~~~~~\ep =1,
\eeq
where $r$ is the lattice geodesic distance (the 
link distance) and   $l$ is the lattice length version of $L$. It is 
defined in the following way: for a given triangulation 
let $i$ be a given vertex and let 
$j_1, \ldots,j_n$ be the set of  vertices located a link distance $r$
from vertex $i$. The vertices $j_1,\ldots,j_n$ can be divided in
connected \emph{maximal} subsets such that $j_\a$ belongs to 
the subset $\{j_{\b_1},\ldots,j_{\b_k}\}$ if it is neighbor to any of 
the $j_\b$'s and not neighbor to a $j_\a$ in 
$\{j_1,\ldots,j_n\}/\{j_{\b_1},\ldots,j_{\b_k}\}$. 
For such a maximal subset we define $l \equiv k$. The loop length 
distribution $\rho_N(r,l)$ at the discretized level measures 
the average number of such connected boundaries of length $l$ corresponding
to the geodesic radius $r$ for the ensemble of triangulations with $N$
triangles. Corresponding to \rf{*in32} we have 
\beq{iii8}
\la \, l^n(r)\,\ra_N = \sum_l l^n \, \rho_N(r,l),
\eeq
while the scaling hypothesis \rf{*in37a} becomes
\beq{iii9a}
\la \, l^n\, (r) \ra_N = N^{2n/d_h} F_n(x),
\eeq
where
\beq{iii10}
F_n(x) \sim x^{2n} ~~~~{\rm for} ~~~x \ll 1.
\eeq
As already mentioned, we will provide evidence that 
the loop length distribution in the limit $N\to\infty$, 
even after gravity is coupled to matter, has the form 
\beq{iii10a}
\rho_{N\!=\!\infty} (r,l) = \frac{1}{r^2}\; \hat{\rho} (y),
~~~~y = \frac{l}{r^2}.
\eeq
Numerical simulations of pure gravity give good agreement with 
$\hat{\rho}(y)$ given by \rf{*in35} \cite{fty}.
Here we will be interested in a determination of $\rho_V(R,L)$ for
the critical Ising and the critical three-states Potts model 
coupled to quantum gravity.

\section{Numerical Simulation}

The numerical simulations are performed as follows:  The Monte Carlo
updating of the triangulations is performed by the so{--}called flip
algorithm and the spins are updated by standard cluster
algorithms. The flips are organised in ``sweeps'' which consist of
approximately $N_L$ {\it accepted} flips where $N_L$ is the number of
links of the triangulated surface.  After a sweep we update the spin
system.  All this is by now standard and we refer to
\cite{bj,atw} for details about the actions or Monte Carlo
procedures. We use the high quality random number generator RANLUX
\cite{mlfj} whose excellent statistical properties are due to its
close relation to the Kolmogorov K{--}system originally proposed by
Savvidy et.al.\cite{ssa} in 1986.

All our runs for the Ising and three{--}states Potts model were
made at the exactly known infinite volume critical temperatures
$\beta_c$ \cite{bk,daul}. The sizes and the number of sweeps we use
are different depending on the observables that we measure. The
largest amount of statistics we gathered were for measuring
$n_1(r;N)$ and $n_\vph(r;N)$ for the Ising and
three{--}states Potts models coupled to gravity, were we performed
$1.7-5.0\times 10^6$ sweeps on surfaces with $16000${--}$128000$
triangles. We also report results obtained on surfaces with $256000$
triangles with a smaller number of sweeps ($\sim 0.8\times 10^6$). For
the moments $\vev{l^n}$ we needed much less statistics: We performed
$3.0-6.0\times 10^5$ sweeps for the $16000${--}$64000$ lattices and
$3.5-7.0\times 10^4$ for the $128000$ lattices. Although less
statistics was necessary in the latter case, the computer effort
needed for the measurements is quite significant compared to that of
measuring $n_1(r;N)$ and $n_\vph(r;N)$, especially for the largest
lattices.

We can extract the Hausdorff dimension from the scaling hypothesis 
\rf{*in30}-\rf{*in30a}:
\beq{iii3} 
n_1 (r;N) = N^{1-1/d_h} F_1(x),~~~~~~n_\vph (r;N) =
N^{1-\D-1/d_h} F_\vph(x) 
\eeq 
as was done in \cite{syracuse,ajw}.  
The analysis of Eq.\ \rf{iii3} is performed by ``collapsing''
$n_{1,\vph}(r;N)$ for a given number of lattice sizes. A fit to
$p_n(x){\e}^{-\alpha x}$, thought of as an interpolating function,
is performed for given $(d_h,a)$ and a value for the $\chi^2(d_h,a)$
is obtained. $p_n(x)$ is a polynomial of $x$ of order $n$. $n$ is
chosen large enough to capture the functional form of
$n_{1,\vph}(r;N)$ by checking that $\chi^2(d_h,a)$ does not depend on
$n$ for a range of $n$ and small enough in order to leave enough
degrees of freedom.  The errors which we
use in the determination of $\chi^2(d_h,a)$ are computed by binning
our data. We refer the reader to the figures in \cite{syracuse,ajw} in
order to appreciate pictorially the impressively good scaling that
$n_{1,\vph}(r;N)$ exhibit in the simulations.

In Fig.~\ref{f:1} and Fig.~\ref{f:2} we show the results from the analysis
of $n_{1,\vph}(r;N)$. We see that determining $a$ is crucial for
extracting $d_h$. In these figures we show $d_h(a)$ given by the value
of $d_h$ which minimises $\chi^2(d_h,a)$ for fixed $a$. The errors are
computed by the interval of $d_h$ which changes $\chi^2(a)\to {\rm
max}\{2,2\chi^2(a)\}$ where $\chi^2(a)={\rm min}_{d_h}\{\chi^2(d_h,a)\}$. As we
can see, the value of $d_h$ changes considerably with $a$ so we need
to determine the range of its acceptable values, since $a=0$ is by no
means a special choice in our consideration. We do this by minimising
$\chi^2(a)$ which gives the results shown in Table~\ref{t:0} and
Table~\ref{t:0b}. The errors quoted are computed by considering the
interval of $a$ which changes $\chi^2_{min}\to {\rm
max}\{2,2\chi^2_{min}\}$ where $\chi^2_{min}={\rm
min}_{a}\{\chi^2(a)\}$. Notice that $a=0$ is not significantly far
from the optimal choice of $a$. This explains the good quality of the
results reported in \cite{syracuse,aamt} in the case of link
distance. This is not true if one uses the triangle distance for the
lattice sizes we consider here (we expect $a$ to become numerically
less significant for large {\it linear} size of the system).

$d_h$ can also be determined from the small $x$ behaviour
\beq{iii4}
F_1(x)  \sim x^{d_h-1},   ~~~~~~ F_\vph(x) \sim x^{d_h(1-\Delta)-1}\, ,
\eeq
and
\bes
\label{iii5}
\slabel{iii5a}
\frac{d \log F_1(x)}   { d \log x} &=& d_h - 1 + x^k +\ldots \, ,\\
\slabel{iii5b}
\frac{d \log F_\vph(x)}{ d \log x} &=& d_h(1-\Delta)-1 + x^l +\ldots \, .
\ees
The calculation of the logarithmic derivative is performed by using a
9 point Savitzky{--}Golay smoothing filter \cite{sg} with a ${\rm
8}^{\rm th}$ order interpolating polynomial.  The errors are computed
by binning our data. The use of the filter improves the computation of
the derivatives, especially near the origin.  The analysis of the
logarithmic derivatives provides an excellent pictorial way for
realizing the scaling given by Eq.~\rf{iii4} near the
origin. Fig.~\ref{f:3} shows no such scaling if $a=0.0$. Similar lack
of scaling for $a=0.0$ is observed for all correlation functions we
analyzed near the origin. There is an optimal value of $a$, however,
where scaling becomes manifest. This is shown in Fig.~\ref{f:3} and
Fig.~\ref{f:3a}. In order to determine the optimal $a$ we use direct
fits to Eq.\rf{iii4}. The results are shown in Table~\ref{t:2} and
Table~\ref{t:3}\footnote{We should warn the reader that the values of
$\chi^2$ reported for all fits are to be strictly compared only for
same lattice sizes since the amount of statistics varies between
lattice sizes.}. In order for the correlation functions to have a
``smooth'' continuum limit \cite{yw} it is very important that the
value of $d_h$ extracted from those fits is the same as the one
extracted from the analysis of Eq.~\rf{iii3} since in principle the
two values can be different. For $n_1(r;N)$, $d_h$ extracted from
finite size scaling variable $x$ corresponds to $d_H$ of
Eq.~\rf{*global1} and $d_h$ extracted from the small distance
behaviour corresponds to $d_h$ of Eq.~\rf{*in15}. In the case of
$n_\vph(r;N)$, agreement between the two values implies also the existence
of a diverging correlation length for the matter fields
\cite{aamt,aamt2}, a fundamental assumption for scaling in critical
phenomena. For matter systems coupled to gravity we could find the
situation where, despite the fact that we have a continuous phase
transition, the scale associated with the geometry diverges whereas
the scale associated with matter does not, as is the case of many
Ising spins coupled to two{--}dimensional quantum gravity
\cite{ha}. From our results we conclude that we do obtain the same
scaling exponents in Eq.~\rf{iii3} and Eq.~\rf{iii4}, the small
differences being attributed to finite size effects. We also observe
that by introducing the shift $a$, scaling at small $x$ with
reasonable values for $d_h$ appears from the analysis of much smaller
lattices than it was thought it would be necessary before. Our results
on the largest lattices are, however, necessary in order to gain
confidence that the fits are stable with respect to changing the
points that one includes in the fits. $d_h$ decreases slightly by
removing points from the origin, but the fits are more stable for the
largest lattice that we use.

At this point we would like to present our results for the ordinary
spin{--}spin correlation function $G_\vph(r;N)$. It is a pleasent
surprise that it exhibits excellent scaling properties, in fact
better than the ``unnormalized'' function $n_\vph(r;N)$. The reason
for the improved quality is presumable that some correlated
fluctuations in spin and geometry are cancelled in $G_\vph(r;N)$.  Our
results for the value of $d_h$ obtained from $n_\vph(r;N)$ are
confirmed and give us further confidence on the existence of a
diverging correlation length for the matter fields.  Recall that we
expect the scaling
\beq{iii7}
G_\vph(r;N)\sim N^{-\Delta} g_\vph(x)\, ,
\eeq
where
\beq{iii7a}
g_\vph(x) \sim  x^{-d_h \Delta}\, ,\qquad x\ll 1\, .
\eeq
In Fig.~\ref{f:4} we see that the scaling \rf{iii7} holds very well if
one chooses the appropriate value of $a$. $g_\vph(x)$ has a stronger
dependence on $a$ near the origin than $F_{1,\vph}(x)$ and we do not
obtain good scaling at $a=0$. The value of $a$ chosen in the plots is
taken from the fits to Eq.~\rf{iii7a}. The results of the latter are
shown in Table~\ref{t:4}. As one can see from the plot of the
logarithmic derivative of $g_\vph(x)$ in Fig.~\rf{f:5}, the fits are
stable over a wider range than those of Eq.~\rf{iii4} for
$n_\vph(r;N)$ and we get and excellent agreement with the value of
$d_h$ we have obtained so far. In Table~\ref{t:4} we show our results
for different cuts for the range in $r$: We compare the same range
used in Table~\ref{t:3} where applicable, a typical point in the
region of stability and finally the point where $\chi^2$ becomes of
order $1$.  This result further supports the statement that the
correlation length for the spin{--}spin correlation function diverges
at the critical temperature.

The results for the Hausdorff dimension are further tested 
by measuring the loop length distribution $\rho_N(r,l)$ and calculating
\beq{iii9}
\vev{l^n(r)}_N \sim N^{\frac{2 n}{d_h}} F_n(x)\, , \qquad n=2,3,4\, ,
\eeq
where 
\beq{iii10b}
F_n(x) \sim x^{2n}\, ,\qquad x\ll 1\, .
\eeq
We have analyzed Eq.~\rf{iii9} in a similar way we have analyzed
$n_{1,\vph}(r;N)$ by collapsing the distributions. In Fig.~\ref{f:6}
we show the collapsed $\vev{l^2(r)}_N$ function and we observe that
scaling holds very convincingly. Quite similar plots can be obtained
for $\vev{l^{3,4}(r)}_N$. We observe that the extracted $d_h(a)$ has a
very weak dependence on the shift $a$ contrary to what we found for
$\vev{l^1(r)}_N$ before. This is shown in Fig.~\ref{f:7} for the Ising
model. Similar graphs can be obtained for all models and moments and
we show representative values of $d_h(a)$ in Table~\ref{t:1}. In
Table~\ref{t:1a} we show the best value for the shift $a$ and the
corresponding value $d_h(a)$. This is slightly different than the
procedure we followed in the analysis of $n_{1,\vph}(r;N)$, since now
$d_h(a)$ is not invertible. In Fig.~\ref{f:5} we show that the scaling
of Eq.~\ref{iii10} holds very well near the $r$ origin.

Finally in Fig.~\ref{f:9} {--} Fig.~\ref{f:11} we show our results for
the loop{--}length distribution function $\rho_N(r,l)$. In
Fig.~\ref{f:9} we show the pure gravity measurements together with a
fit to Eq.~\rf{*in35}. In the fit we simply rescale $y$ and
$\rho_N(r,l)r^2$, but we otherwise keep the same coefficients as in
Eq.~\rf{*in35}. The data is consistent with the theoretical prediction,
as was found before in \cite{fty}, but unfortunately it has no
predictive power to determine convincingly the terms in Eq.~\rf{*in35}
for $c\not= 0$. We are able to check, however, some of the scaling
properties of $\rho_N(r,l)r^2$: The part of the curve corresponding to
the continuum behaviour is independent of $N$ and that
$\rho_N(r,l)(r/N^{1/d_h})^2= N^0 F_\rho(y,L/N^{2/d_h})$ for all values
of $y$ and $L/N^{2/d_h}$. This is shown in Fig.~\ref{f:12} for the
three{--}states Potts model. It is also curious that in the range in $y$
where we observe continuum behaviour, $\rho_N(r,l)r^2$ seems to be
independent of $m=2,3,5$. In this range the data points in
Fig.~\ref{f:10} are within statistical error on top of each other. One
could be tempted to conjecture that the fractal properties of
space{--}time for unitary theories with $0\leq c < 1$ are independent
of $c$. In the same figure, however, we observe that the finite size
corrections do depent on $c$ and it is not so clear which is the
borderline in the range in $y$ where we observe continuum physics and
where finite size effects are important. In contrast, for the $c=-2$
model simulated in
\cite{dgi,dgi2} we see a clear difference from our data, reflecting
the fact that in this model the fractal dimension of space{--}time is
$3.58(4)$. In Fig.~\ref{f:11} we compare our data with that of
\cite{dgi,dgi2}.

\section{Discussion}

In this work we have measured the Hausdorff dimension $d_h$ for $0\leq
c<1$ matter coupled to two{--}dimensional quantum gravity using
various scaling arguments. In the spirit of \cite{ajw}, we introduced
the shift $a$ in the investigation, which as a finite size correction,
improves dramatically the scaling of correlation functions for small
geodesic distances $r$ giving consistent results for $d_h$. For finite
size scaling, the shift must be included in the analysis, even when we
use the link distance in the definition of correlation functions, and
can be used to estimate the systematic errors introduced by finite
size effects. Its effect is greatest in the case of the two point
functions $n_{1,\vph}(r;N)$ and $G_\vph(r;N)$. By studying the scaling
behaviour of the spin{--}spin correlation functions we verified and
extended the results of \cite{aamt,aamt2} on the existence of a
diverging correlation length for matter. We also verified the scaling
properties of the moments $\vev{L^n}$ for pure gravity predicted in
\cite{transfer,dgi} and found that similar scaling holds for $0<c<1$
matter coupled to gravity. We measured the loop{--}length distribution
function $\rho_N(r,l)$ and showed that in the continuum limit
$\rho_\infty(R,L)$ is not or very little affected by the back reaction
of $0<c<1$ matter to gravity.

The results of our measurements of $d_h$ are consistent with the
earlier observations in \cite{syracuse,ajw} that the presence of
$0<c<1$ matter has no or very small effect on $d_h$. This is in
contradiction with the analytic result $d_h = 2m$, as was mentioned in
the introduction. It has been argued that the reason for the apparent
contradiction is that a large Hausdorff dimension $d_h=2m$,
$m=3,5,\ldots$ implies a very small linear extension $N^{1/d_h}$ for
the range of $N$ accessible in the numerical simulations.  However, if
this argument was correct it would be very hard to understand how one
can measure with excellent precision the correct {\it KPZ} exponents
for the Ising and three-states Potts models from the integrated
correlators.  Moreover the correct scaling of the correlation
functions defined in terms of geodesic distance provide even stronger
evidence that we see the correct coupling of matter to the geometry in
the simulations, including the fractal properties of the metric.  If
the linear size of the systems were much too small one should not be
able to observe the correct critical behavior of conformal matter
fields coupled to quantum gravity. Moreover, as it was already
mentioned, it is known
\cite{kketal,dgi,dgi2} that $d_h = 2m$ is inconsistent with numerical
simulations on the $c=-2$ model coupled to gravity, which corresponds 
to $m=1$, i.e.\ $d_h=2$. In this case the simulated systems have a quite
large linear size.

As it was mentioned in the introduction the alternative prediction
coming from scaling arguments for the diffusion equation in Liouville
theory,
\beq{iv2}
d_h =     2\times\frac{\sqrt{25-c}+\sqrt{49-c}}{\sqrt{25-c}+\sqrt{1-c}},
\eeq
is in excellent agreement with the $c=-2$ simulations, and it agrees 
with the rigorously established result $d_h=4$ for $c=0$.

{\it Our simulations are consistent with the $d_h=4$ conjecture 
\cite{syracuse,ajw} for Ising
and three{--}states Potts model, especially when looking at
correlators not involving the matter fields.} Although the reader can
draw her/his own conclusions from our measurements, we included a
summary of our results in Table~\ref{t:5} where we display the most
probable range for $d_h$ as measured by the different scaling
arguments we used in this article. We observe that the values of $d_h$
coming from correlation functions involving matter fields are
consistently higher. The errors are big enough to include $d_h=4$
within 1{--}1.5$\sigma$ and one can argue that finite size effects are
bigger in this case: As pointed out in \cite{syracuse}, the scaling
behaviour of the spin{--}spin correlation function $n_\phi(r;N)$ is
the difference $f_1(r)-1/(q-1)f_2(r)$ of the correlators $f_1(r)$ and
$f_2(r)$ which count the number of like and different spins at
distance $r$. They both scale identically as $n_1(r;N)$ and the
scaling behaviour of $n_\vph(r;N)$ is obtained by exact cancellation
of the leading terms. In favour with the $d_h=4$ conjecture is our
result that $\rho_N(r,l)$ seems to be independent of $m$ in the region
where it exhibits continuum behaviour.

{\it One, however, cannot claim that our simulations exclude the prediction
of Eq.~\rf{iv2}}. Except for $n_1(r;N)$, one can see that $d_h$
increases slightly with $m$, although the signal is not clear enough
to allow for a clear distinction. There exist the possibility that the larger
values of $d_h$ obtained from $n_\phi(r;N)$ and
$G_\phi(r;N)$ are a true signal. Firstly, it {\it is} of course interesting 
that the central values of $d_h$ actually agree with Eq.\ \rf{iv2}, although
with large error bars. Secondly,  one cannot
exclude with certainty the possibility that simulations on larger
lattices will not shift slightly the values of $d_h$ extracted from the 
other observables to the ones of Eq.~\rf{iv2}. In this connection one 
should bear in mind that the finite size effects for the Ising and 
the three-states Potts model are larger than for for pure gravity, as 
shown in \cite{dhy}. Nevertheless one can hope that future 
simulations might decrease systematic errors to the point that one 
will be able to reject convincingly at least one of the above predictions.

As we  noted in the introduction, the Hausdorff dimension comes from
the cutoff dependence as $L\to 0$ of the integral \rf{*in36}. 
The cutoff dependence disappears in the integrals \rf{*in37} for $n >1$
and one obtains the scaling \rf{*in37a}-\rf{*in37b} which seems 
very well satisfied also in the case of the Ising and three-states Potts model
coupled to gravity, as well as for $c=-2$! Therefore, in order to be able
to fully understand the concept of the Hausdorff dimension for
$c\not=0$, one needs to provide an explanation of expressions like
Eq.\rf{*in34} and \rf{*in35} for the loop length distribution function 
$\rho_V(R,L)$, but with different powers of $y$, also for $c \neq 0$.

\subsection*{Acknowledgments}
K.A. would like to acknowledge interesting 
discussions with Gudmar Thorleifsson. 
J.A. acknowledges the support of the Professor Visitante Iberdrola
grant and the hospitality at the University of Barcelona, where part
of this work was done.

%% file: f_tab.tex
\begin{table}[ht]
\begin{center}
\begin{tabular}{|c | l c | r c l | }
\hline
$m$&    \multicolumn{1}{|c }{$d_h$} & $a$ &  \multicolumn{3}{|c|}{$N_T$}\\
\hline
2  &   4.05(8)    &    0.60(20)    & 8000  & {--} & 64000\\
3  &   4.11(10)   &    0.48(28)    & 16000 & {--} & 128000\\
5  &   4.01(9)    &    0.15(26)    & 16000 & {--} & 128000\\
2  &   4.05(15)   &    0.70(60)    & 16000 & {--} & 64000\\
3  &   4.11(11)   &    0.50(40)    & 32000 & {--} & 128000\\
5  &   3.98(15)   &    0.00(60)    & 32000 & {--} & 128000\\
\hline
\end{tabular}
\end{center}
\caption{The Hausdorff dimension $d_h$ as determined from
collapsing the $n_1(r;N)$ correlation functions for pure gravity
($m=2$), Ising ($m=3$) and three{--}states Potts model ($m=5$) coupled
to gravity.}
\label{t:0}
\end{table}

\begin{table}[ht]
\begin{center}
\begin{tabular}{|c | l c | r c l | }
\hline
$m$&    \multicolumn{1}{|c }{$d_h$} & $a$ &  \multicolumn{3}{|c|}{$N_T$}\\
\hline
3  &  4.28(17)  &   0.60(30)  &  8000 & {--} & 128000\\
5  &  4.46(33)  &   0.53(51)  &  8000 & {--} & 128000\\
3  &  4.26(26)  &   0.56(48)  & 16000 & {--} & 128000\\
5  &  4.45(40)  &   0.55(95)  & 16000 & {--} & 128000\\
\hline
\end{tabular}
\end{center}
\caption{The Hausdorff dimension $d_h$ as determined from
collapsing the $n_\varphi(r;N)$ correlation functions for pure gravity
Ising ($m=3$) and three{--}states Potts model ($m=5$) coupled to
gravity.}
\label{t:0b}
\end{table}

\begin{table}[ht]
\begin{center}
\begin{tabular}{|c|c c| c c|c c|}\hline
    & \multicolumn{6}{| c |}{ $d_h$} \\
\cline{2-7} 
$n$ & \multicolumn{2}{| c |}{$m=2$} 
    & \multicolumn{2}{| c |}{$m=3$} 
    & \multicolumn{2}{| c |}{$m=5$}  \\
    &   $a=0.00$     &  $a=0.25$     
    &   $a=0.00$     &  $a=0.25$     
    &   $a=0.00$     &  $a=0.25$     \\
\hline
2   & 3.88(3) & 3.90(3)& 3.99(4)& 4.01(4)& 4.07(3)& 4.10(3) \\
3   & 3.94(3) & 3.96(3)& 4.06(4)& 4.09(4)& 4.15(4)& 4.17(4) \\
4   & 3.95(3) & 3.97(3)& 4.08(4)& 4.10(4)& 4.16(4)& 4.18(4) \\
\hline
\end{tabular}
\end{center}
\caption{The Hausdorff dimension $d_h(a)$ as determined from
collapsing the $\vev{l^n(r)}_N$ distributions for $N_T=16000$, $32000$ and
$64000$ for pure gravity ($m=2$), Ising ($m=3$) and three{--}states
Potts model ($m=5$) coupled to gravity.}
\label{t:1}
\end{table}
 

\begin{table}[ht]
\begin{center}
\begin{tabular}{|c | l l | l l| l l |}
\hline
$n$ &  \multicolumn{2}{c|}{$m=2$} & 
       \multicolumn{2}{c|}{$m=3$} & 
       \multicolumn{2}{c|}{$m=5$} \\
    &  \multicolumn{1}{c}{$d_h$} &  \multicolumn{1}{c|}{$a$} &
       \multicolumn{1}{c}{$d_h$} &  \multicolumn{1}{c|}{$a$} &
       \multicolumn{1}{c}{$d_h$} &  \multicolumn{1}{c|}{$a$} \\
\hline
 \multicolumn{7}{|c|}{$N_T=16000${--}$64000$}\\
\hline
2  & 3.88(3) & 0.00(15) & 4.00(4) & 0.0(3)   & 4.08(3) & 0.0(3)\\
3  & 3.94(3) & 0.0(2)   & 4.08(4) & 0.0(4)   & 4.16(4) & 0.1(3)\\
4  & 3.95(3) & 0.00(15) & 4.10(4) & 0.0(5)   & 4.17(4) & 0.1(4)\\
\hline
 \multicolumn{7}{|c|}{$N_T=4000${--}$64000$}\\
\hline
2  & 3.86(2) & 0.00(10) & 3.98(4) & 0.05(10) & 4.07(4) & 0.12(10)\\
3  & 3.92(2) & 0.05(10) & 4.05(4) & 0.07(18) & 4.16(3) & 0.15(15)\\ 
4  & 3.92(2) & 0.05(10) & 4.06(5) & 0.10(18) & 4.17(3) & 0.13(15)\\
\hline

\end{tabular}
\end{center}
\caption{The Hausdorff dimension $d_h(a)$ as determined from collapsing
the $\vev{l^n(r)}_N$ correlation functions for pure gravity ($m=2$),
Ising ($m=3$) and three{--}states Potts model ($m=5$) coupled to
gravity. The best value for the shift $a$ and the corresponding
$d_h(a)$ is recorded.}
\label{t:1a}
\end{table}

\begin{table}[ht]
\begin{center}
\begin{tabular}{|c r |  l l l  |c c c|}\hline
$m$& \multicolumn{1}{c |}{$N_T$}& 
     \multicolumn{1}{c }{$d_h$} & 
     \multicolumn{1}{c }{$a$}   & 
     \multicolumn{1}{c|}{$C$}
      & $\chi^2$& $r_{min}$& $r_{max}$ \\
\hline
3& 256000& 4.081(4)& 0.514(4)& 1.049(9)& 3.7& 1& 6\\
 & 128000& 4.098(7)& 0.529(5)& 1.01(1) & 2.1& 1& 5\\
 &  64000& 4.080(5)& 0.517(4)& 1.06(1) & 6.1& 1& 4\\
 &  32000& 4.041(6)& 0.492(5)& 1.11(1) & 7.4& 1& 4\\
 &  16000& 3.969(9)& 0.448(7)& 1.25(2) & 8.6& 1& 4\\
 & 256000& 4.028(7)& 0.447(9)& 1.19(2) & 1.3& 2& 7\\
 & 128000& 3.994(7)& 0.417(9)& 1.27(2) & 1.8& 2& 6\\
 &  64000& 3.961(8)& 0.39(1) & 1.36(3) & 1.7& 2& 5\\
 &  32000& 3.86(1) & 0.30(1) & 1.65(4) & 5.1& 2& 5\\
 &  16000& 3.68(1) & 0.14(2) & 2.33(7) & 6.7& 2& 5\\
\hline
5& 256000& 4.096(5)& 0.482(5)& 1.07(1) & 3.5& 1& 6\\
 & 128000& 4.082(4)& 0.482(5)& 1.07(1) & 7.4& 1& 5\\
 &  64000& 4.087(5)& 0.481(4)& 1.08(1) & 6.9& 1& 4\\
 &  32000& 4.042(7)& 0.453(6)& 1.17(2) & 7.0& 1& 4\\
 &  16000& 3.964(4)& 0.406(4)& 1.32(1) & 42 & 1& 4\\
 & 256000& 4.031(9)& 0.40(1) & 1.25(3) & 1.2& 2& 7\\
 & 128000& 3.990(7)& 0.367(9)& 1.36(2) & 2.6& 2& 6\\
 &  64000& 3.950(9)& 0.34(1) & 1.46(3) & 2.1& 2& 5\\
 &  32000& 3.83(1) & 0.24(1) & 1.82(5) & 4.3& 2& 5\\
 &  16000& 3.627(8)& 0.051(9)& 2.69(5) & 30 & 2& 5\\
\hline
2& 128000& 4.01(1) & 0.542(9)& 1.04(3) & 0.8& 1& 5\\
 &  64000& 4.023(6)& 0.551(5)& 1.06(1) & 3.9& 1& 4\\
 &  32000& 4.01(1) & 0.54(1) & 1.09(3) & 1.6& 1& 4\\
 &  16000& 3.94(2) & 0.49(1) & 1.23(4) & 1.5& 1& 4\\
 & 128000& 3.95(2) & 0.47(2) & 1.25(5) & 0.2& 2& 5\\
 &  64000& 3.920(9)& 0.44(1) & 1.34(4) & 0.8& 2& 5\\
 &  32000& 3.84(2) & 0.35(3) & 1.58(8) & 0.4& 2& 5\\
 &  16000& 3.69(3) & 0.22(3) & 2.1(1)  & 1.8& 2& 5\\
\hline
\end{tabular}
\end{center}
\caption{The results of the fits to Eq.~\rf{iii4} for $n_1(r;N)$ 
for pure gravity ($m=2$), Ising ($m=3$) and three{--}states Potts
($m=5$) model coupled to gravity.}
\label{t:2}
\end{table}

\begin{table}[ht]
\begin{center}
\begin{tabular}{|c r |  c c c | c |c c c|}\hline
$m$& \multicolumn{1}{c |}{$N_T$}& 
      $c$& $a$& $C$& $d_h$& $\chi^2$& $r_{min}$& $r_{max}$ \\
\hline
 3 & 256000& 1.75(2)& 0.46(2)& 0.73(3)& 4.13(3)& 0.26& 1& 6\\
   & 128000& 1.74(2)& 0.43(2)& 0.76(2)& 4.10(2)& 0.56& 1& 5\\
   &  64000& 1.72(2)& 0.42(2)& 0.77(2)& 4.08(3)& 0.44& 1& 4\\
   & 256000& 1.68(3)& 0.30(7)& 0.87(7)& 4.01(5)& 0.26& 2& 7\\
   & 128000& 1.62(3)& 0.21(6)& 0.97(6)& 3.94(4)& 0.74& 2& 6\\
   &  64000& 1.58(3)& 0.16(7)& 1.04(8)& 3.87(5)& 0.21& 2& 5\\
\hline
 5 & 256000& 1.56(2)& 0.35(2)& 0.73(2)& 4.27(3)& 1.15& 1& 6\\
   & 128000& 1.56(1)& 0.36(2)& 0.73(2)& 4.27(2)& 0.81& 1& 5\\
   &  64000& 1.57(2)& 0.38(2)& 0.71(2)& 4.28(3)& 0.67& 1& 4\\
   & 256000& 1.43(3)& 0.07(7)& 0.97(6)& 4.06(5)& 0.46& 2& 7\\
   & 128000& 1.43(3)& 0.08(6)& 0.98(6)& 4.05(5)& 0.85& 2& 6\\
   &  64000& 1.36(3)&-0.03(7)& 1.10(7)& 3.93(5)& 0.88& 2& 5\\
\hline
\end{tabular}
\end{center}
\caption{The results of the fits to Eq.~\rf{iii4} for $n_\varphi(r;N)$ for
Ising ($m=3$) and three{--}states Potts ($m=5$) model coupled to
gravity.}
\label{t:3}
\end{table}

\begin{table}[ht]
\begin{center}
\begin{tabular}{|c r |  c c c | c |c c c|}\hline
$m$& \multicolumn{1}{c |}{$N_T$}& 
      $c$& $a$& $C$& $d_h$& $\chi^2$& $r_{min}$& $r_{max}$ \\
\hline
3  & 256000& 1.32(2)& 0.51(4)& 0.49(2)& 3.97(6)& 0.13& 1& 6\\
   &       & 1.34(1)& 0.53(3)& 0.50(1)& 4.00(4)& 0.15& 1& 9\\
   &       & 1.37(1)& 0.59(2)& 0.54(1)& 4.10(3)& 1.3 & 1& 13\\
   &       & 1.38(2)& 0.66(8)& 0.56(4)& 4.13(7)& 0.14& 2& 11\\
   &       & 1.44(2)& 0.86(7)& 0.66(4)& 4.32(6)& 1.5 & 2& 14\\
   & 128000& 1.35(2)& 0.56(4)& 0.52(2)& 4.03(7)& 0.07 & 1& 5\\
   &       & 1.38(1)& 0.61(3)& 0.55(2)& 4.14(4)& 0.47& 1& 8\\
   &       & 1.40(1)& 0.65(2)& 0.58(1)& 4.20(3)& 1.2 & 1& 10\\
   &       & 1.44(3)& 0.81(8)& 0.65(4)& 4.32(8)& 0.26& 2& 9\\
   &       & 1.51(2)& 1.02(7)& 0.77(4)& 4.53(6)& 1.8 & 2& 12\\
   &  64000& 1.35(2)& 0.54(3)& 0.51(2)& 4.05(5)& 0.15& 1& 5\\
   &       & 1.39(1)& 0.61(2)& 0.55(1)& 4.16(3)& 1.9 & 1& 8\\
\hline
5  & 256000& 1.57(2)& 0.55(3)& 0.47(2)& 3.93(5)& 0.08 & 1& 6\\
   &       & 1.59(1)& 0.59(3)& 0.50(2)& 4.00(3)& 0.54& 1& 9\\
   &       & 1.63(1)& 0.64(2)& 0.53(1)& 4.06(3)& 1.6 & 1& 11\\
   &       & 1.67(3)& 0.79(8)& 0.61(4)& 4.18(7)& 0.53& 2& 10\\
   &       & 1.73(2)& 0.94(7)& 0.70(5)& 4.33(6)& 1.7 & 2& 12\\
   & 128000& 1.55(2)& 0.53(3)& 0.46(2)& 3.86(5)
               & 5.1$\times{\rm 10}^{\rm-3}$& 1& 5\\
   &       & 1.57(1)& 0.57(2)& 0.48(1)& 3.93(3)& 0.67& 1& 8\\
   &       & 1.59(1)& 0.60(2)& 0.50(1)& 3.98(3)& 1.7 & 1& 9\\
   &       & 1.61(3)& 0.68(8)& 0.53(4)& 4.03(7)& 0.42& 2& 8\\
   &       & 1.65(3)& 0.77(7)& 0.58(4)& 4.12(6)& 1.1 & 2& 9\\
   &  64000& 1.55(2)& 0.53(3)& 0.45(2)& 3.87(4)& 0.43& 1& 5\\
   &       & 1.60(1)& 0.61(2)& 0.51(1)& 4.01(3)& 2.9 & 1& 7\\
\hline
\end{tabular}
\end{center}
\caption{The results of the fits to Eq.~\rf{iii7a} for the normalized
spin{--}spin correlation function for Ising ($m=3$) and
three{--}states Potts ($m=5$) model coupled to gravity.}
\label{t:4}
\end{table}
 
\begin{table}[ht]
\begin{center}
\begin{tabular}{|l | l| l| l l|}
\hline
\multicolumn{5}{|c|}{$d_h$}\\
\hline
\multicolumn{1}{|c|}{$m=2$}        & 
\multicolumn{1}{c|}{$m=3$}        & 
\multicolumn{1}{c|}{$m=5$}        & 
\multicolumn{2}{c|}{Method} \\
\hline
4.05(15)     & 4.11(10)     & 4.01(9)      & $n_1(r;N)$ & FSS\\
3.92{--}4.01 & 3.99{--}4.08 & 3.99{--}4.10 & $n_1(r;N)$ & SDS\\
3.85{--}3.98 & 3.96{--}4.14 & 4.05{--}4.20 & $\vev{l^n(r)}_N$& FSS\\
             & 4.28(17)     & 4.46(33)     & $n_\varphi(r;N)$ & FSS\\
             & 3.90{--}4.16 & 4.00{--}4.30 & $n_\varphi(r;N)$ & SDS\\
             & 3.96{--}4.38 & 3.97{--}4.39 & $G_\varphi(r)_N$& SDS\\
\hline
\end{tabular}
\end{center}
\caption{A summary of the results for $d_h$ shown in
Table~\ref{t:1}{--}Table~\ref{t:4}. FSS in the Method column stands
for Finite Size Scaling and SDS for Small Distance Scaling.}
\label{t:5}
\end{table}
 

%% file: f_fig.tex

\begin{figure}[htb]
\centerline{\epsfxsize=4.0in \epsfysize=2.67in \epsfbox{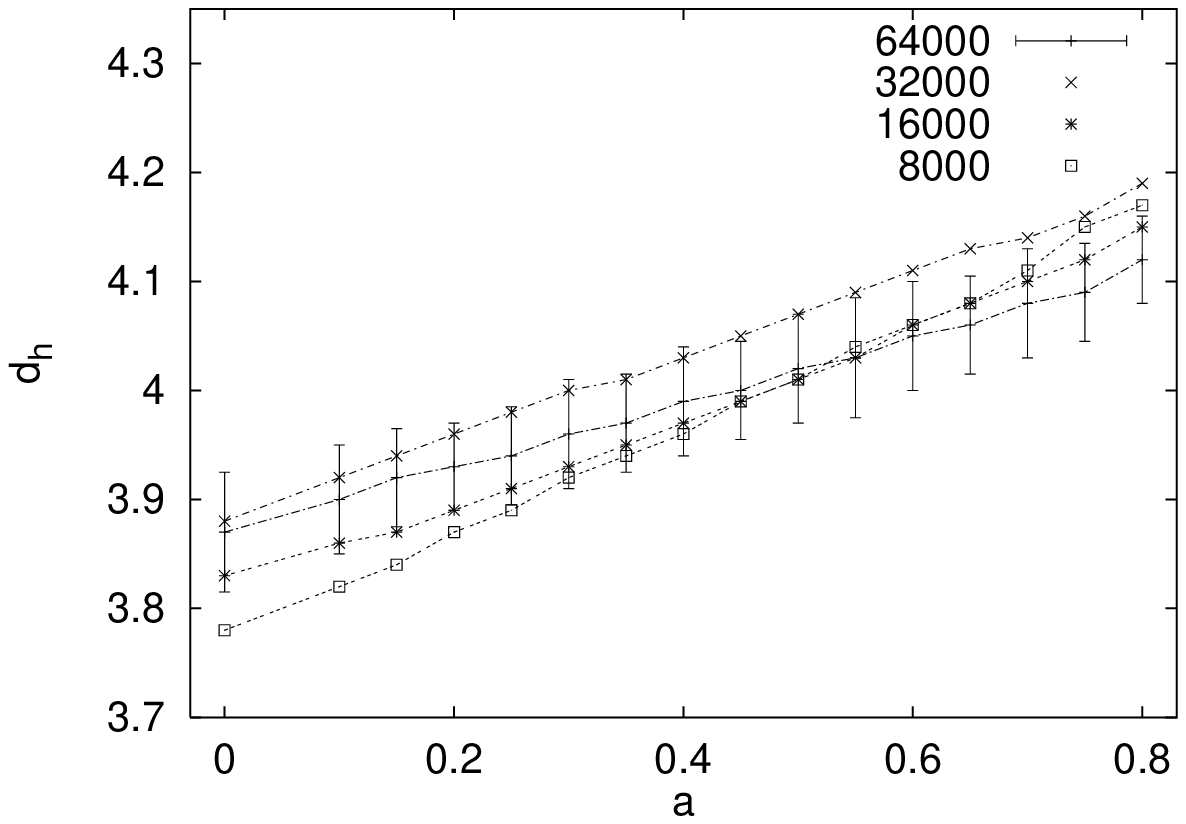}}
\centerline{\epsfxsize=4.0in \epsfysize=2.67in \epsfbox{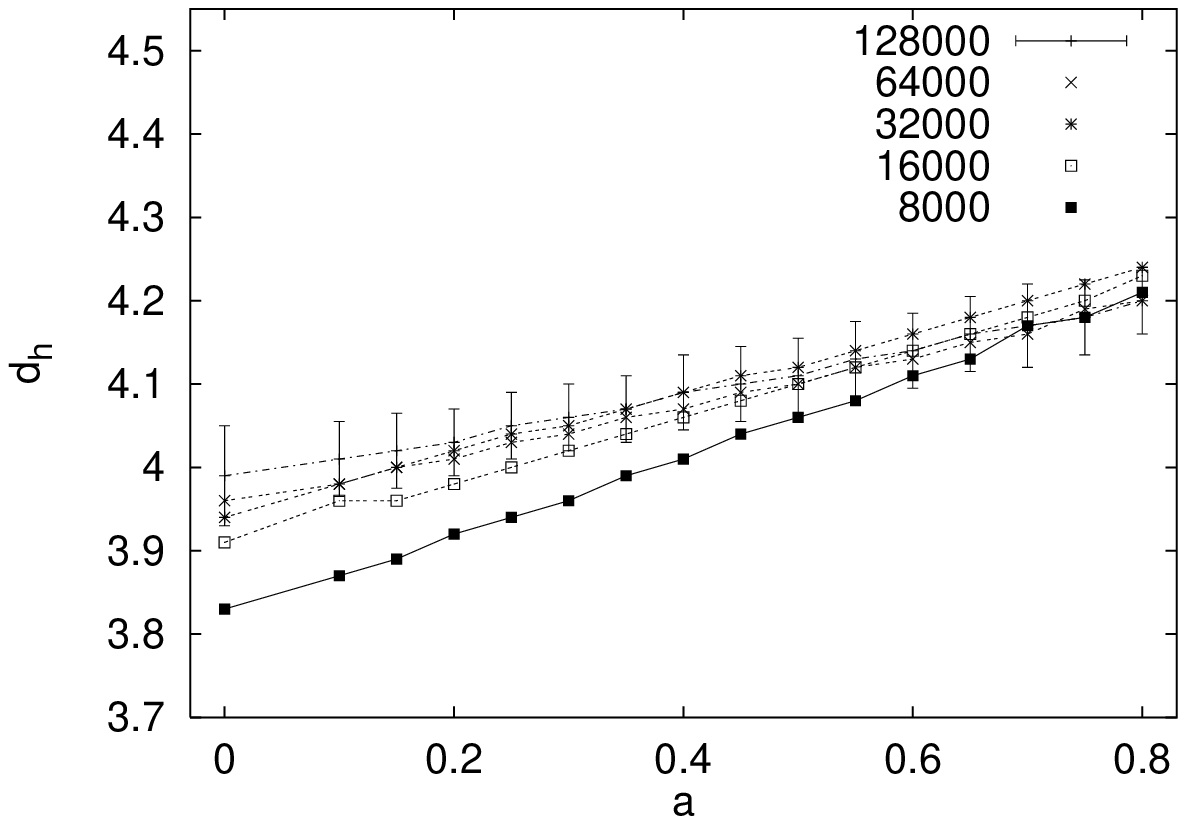}}
\centerline{\epsfxsize=4.0in \epsfysize=2.67in \epsfbox{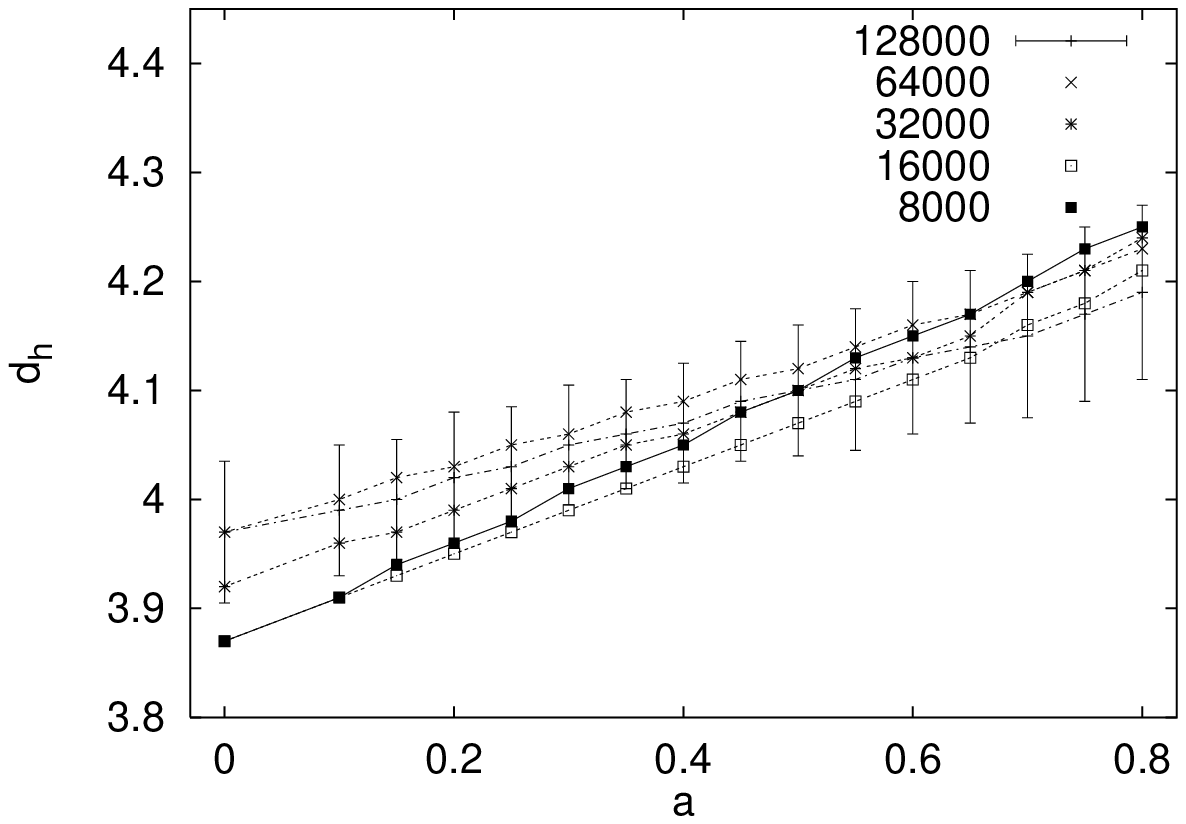}}
\caption{({\it a}) $d_h(a)$ from collapsing $n_1(r;N)$ for $N_T=
8000${--}$64000$ for pure gravity. Data is collapsed in groups of
three lattice sizes and in the graph we indicate the largest of each
group. We show the errors computed from $\chi^2$ only for the largest
lattice in order to simplify the graph. The errors for the smaller
lattices are quite similar.  ({\it b}) Same as in ({\it a}) for the
Ising model for $N_T= 8000${--}$128000$. ({\it c}) Same as in ({\it
b}) for the three{--}states Potts model.}
\label{f:1}
\end{figure}
\clearpage


\begin{figure}[htb]
\centerline{\epsfxsize=4.0in \epsfysize=2.67in \epsfbox{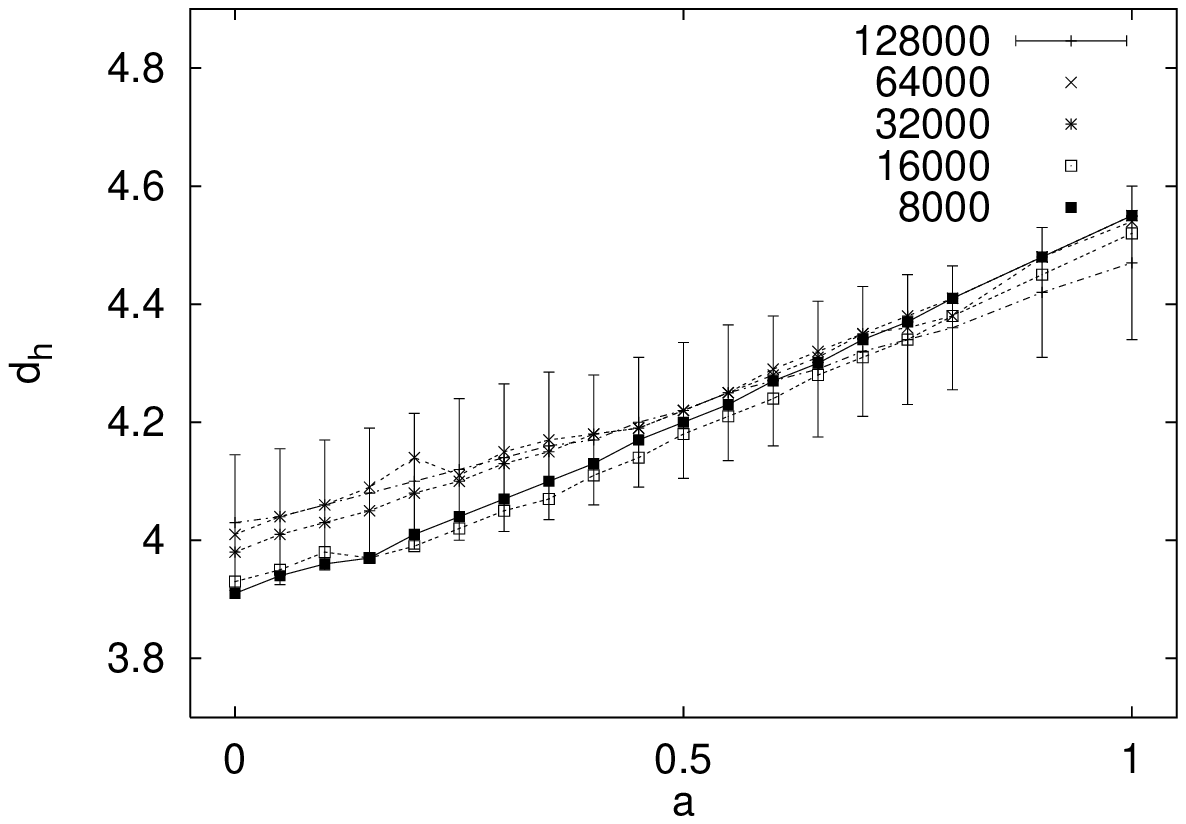}}
\centerline{\epsfxsize=4.0in \epsfysize=2.67in \epsfbox{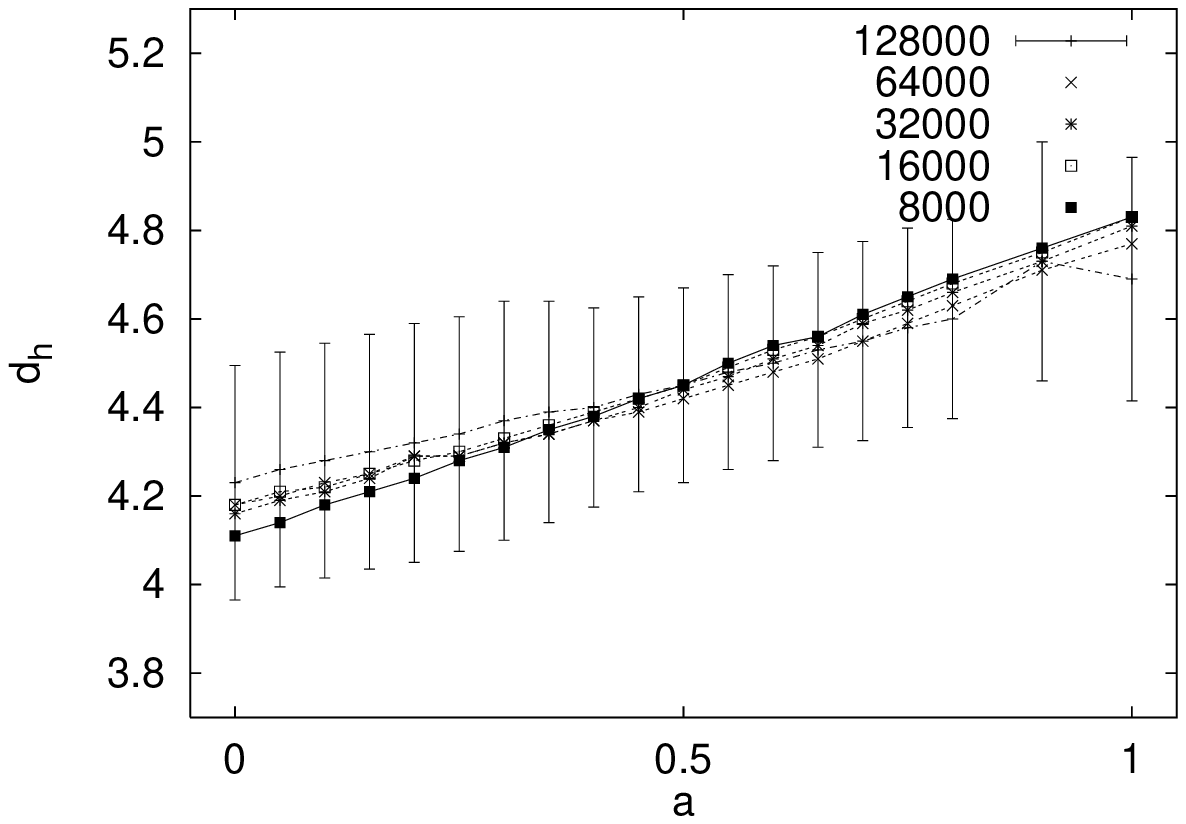}}
\caption{({\it a}) $d_h(a)$ from collapsing $n_\varphi(r;N)$ for $N_T=
8000${--}$128000$ for the Ising model the same way as described in
Fig.~\protect\ref{f:1}. ({\it b})Same as in ({\it a}) for the
three{--}states Potts model.}
\label{f:2}
\end{figure}

\begin{figure}[htb]
\centerline{\epsfxsize=4.0in \epsfysize=2.67in
\epsfbox{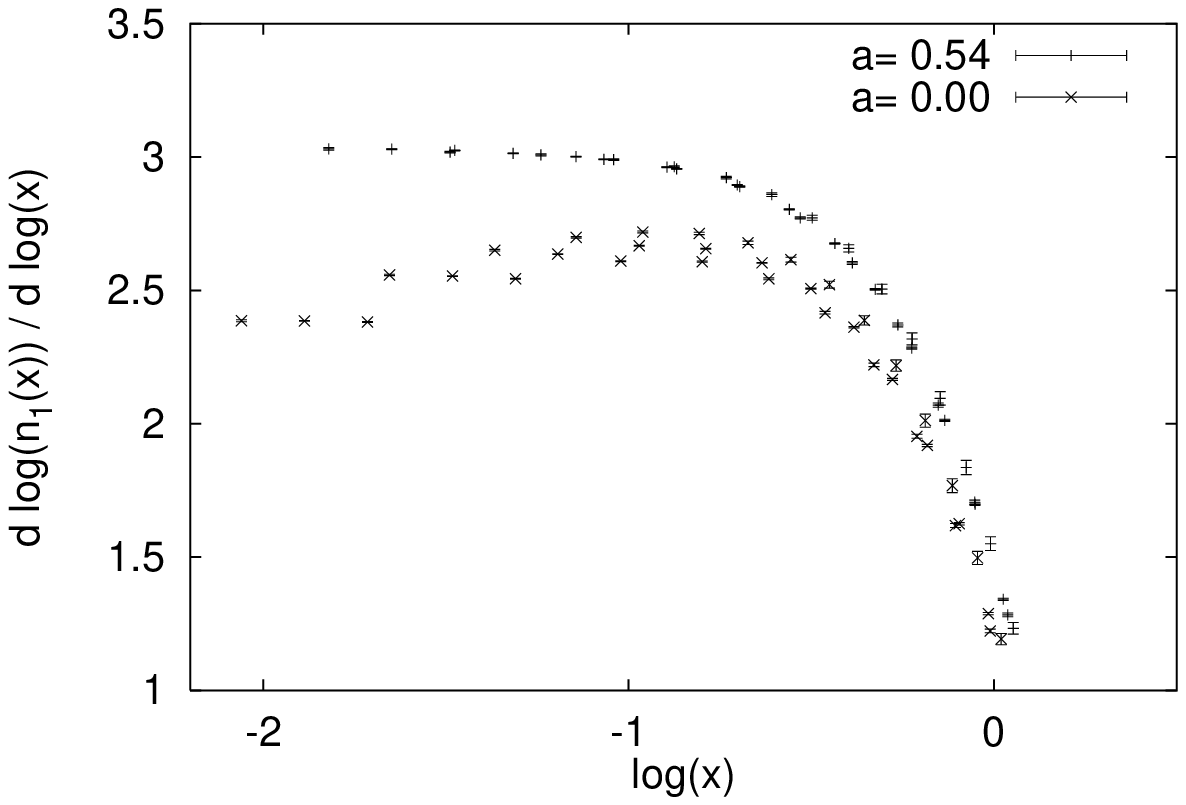}}
\centerline{\epsfxsize=4.0in \epsfysize=2.67in 
\epsfbox{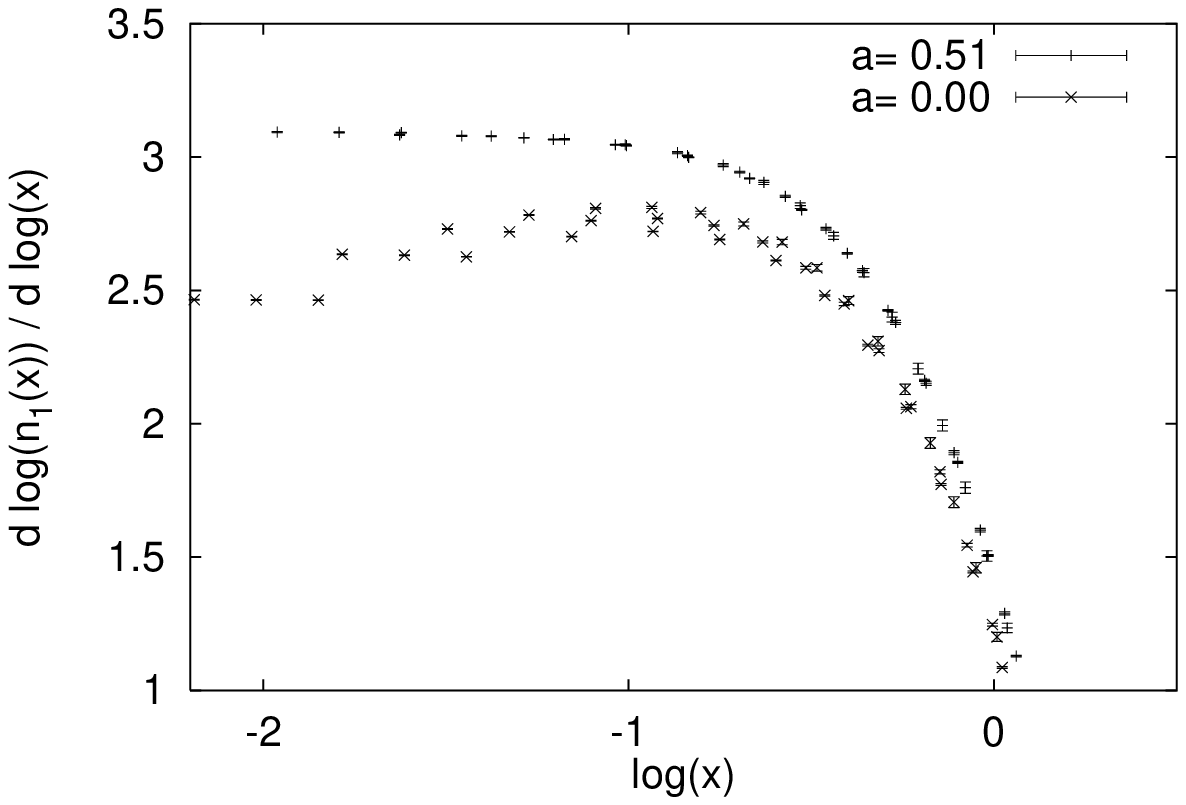}}
\centerline{\epsfxsize=4.0in \epsfysize=2.67in 
\epsfbox{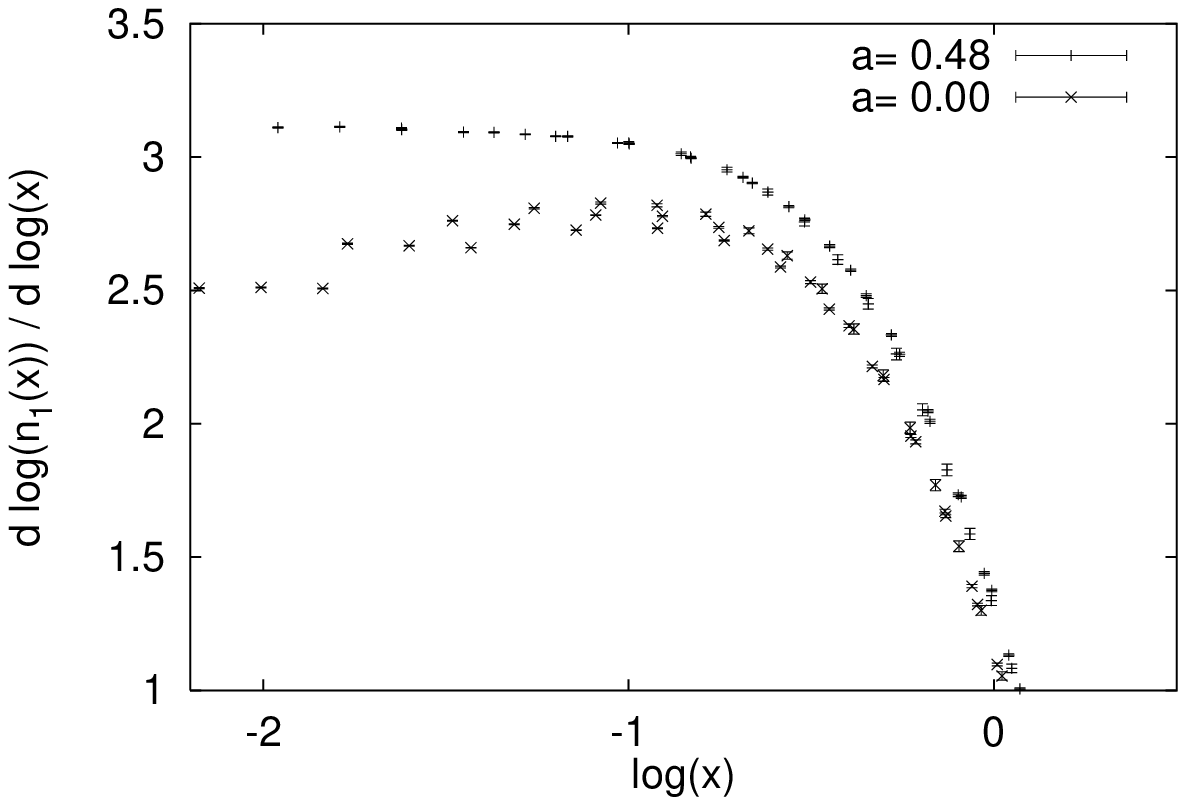}}
\caption{({\it a}) The small $x$ behaviour of the logarithmic derivative 
of $n_1(r;N)$. We use $N_T=32000${--}$128000$, $d_h=4.02$ and
$a=0,0.54$.
({\it b}) Same as in ({\it a}) for the Ising
model coupled to gravity. We plot for $N_T=64000${--}$256000$,
$a=0,0.51$ and $d_h=4.08$.  
({\it c}) Same as in ({\it b}) for the three{--}states
Potts model coupled to gravity where $a=0,0.48$ and $d_h=4.10$.}
\label{f:3}
\end{figure}

\clearpage
\begin{figure}[ht]
\centerline{   \epsfxsize=4.0in \epsfysize=2.67in 
\epsfbox{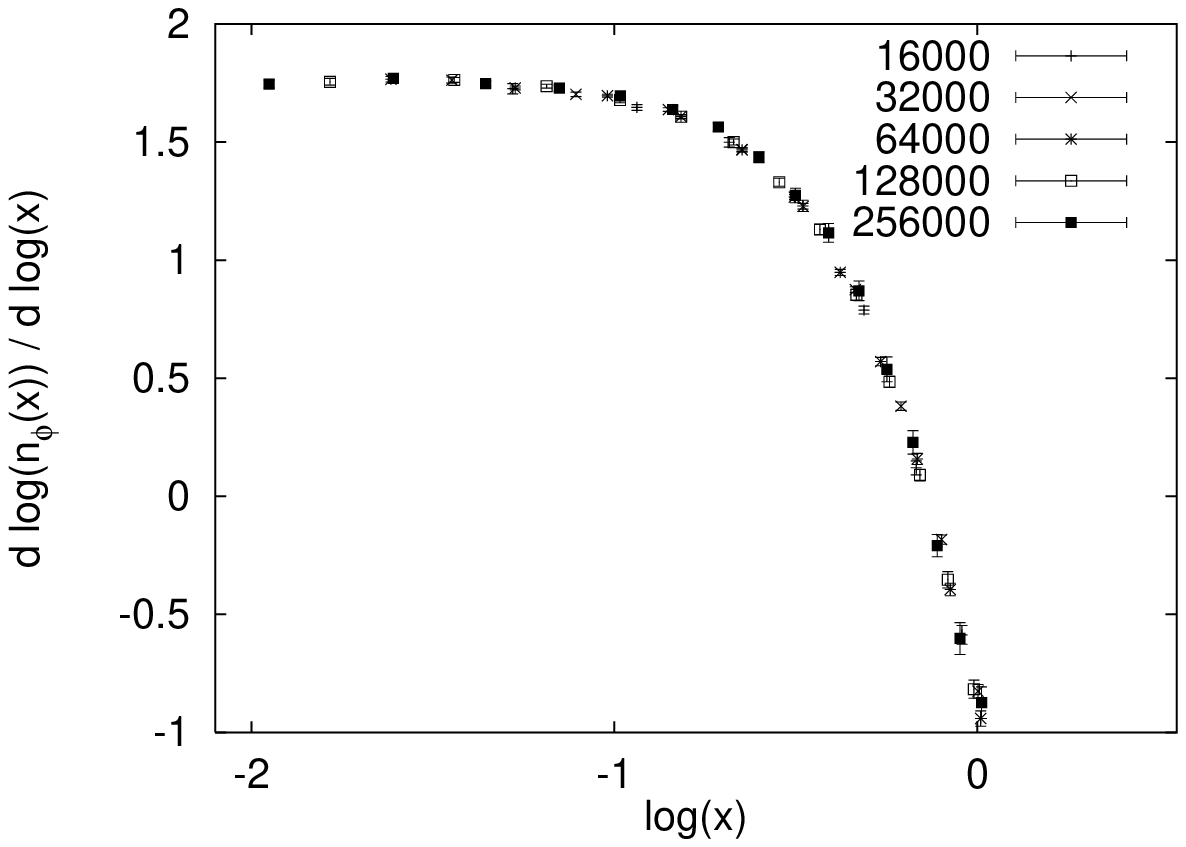}}
\centerline{   \epsfxsize=4.0in \epsfysize=2.67in 
\epsfbox{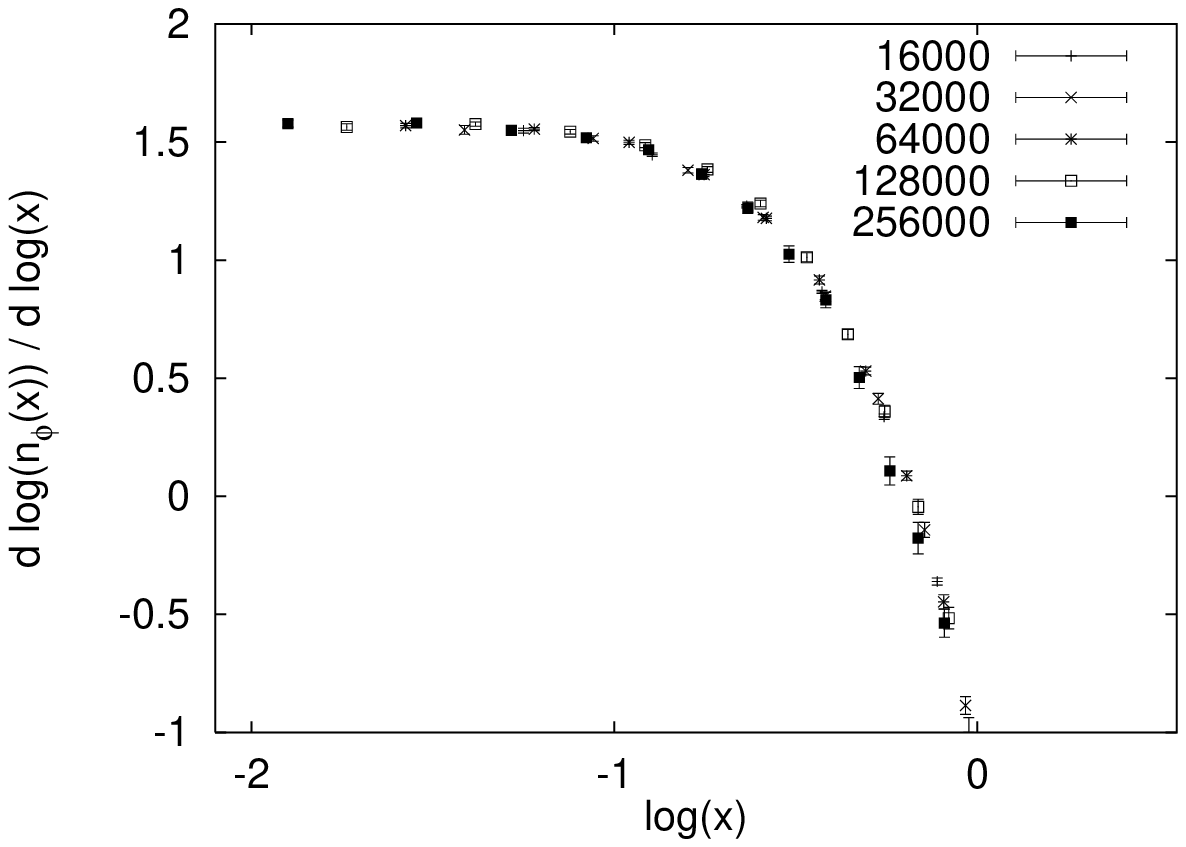}}
\caption{({\it a}) The small $x$ behaviour of the logarithmic
derivative of $n_\varphi(r;N)$ for the Ising model coupled to gravity. We
use $N_T=16000${--}$256000$ and $x$ is obtained by using $d_h=4.13$,
$a=0.45$. 
({\it b}) Same for the three{--}states Potts model coupled to
gravity. We use now $d_h=4.27$ and $a=0.35$.}
\label{f:3a}
\end{figure}

\begin{figure}[htb]
\centerline{   \epsfxsize=4.0in \epsfysize=2.67in 
\epsfbox{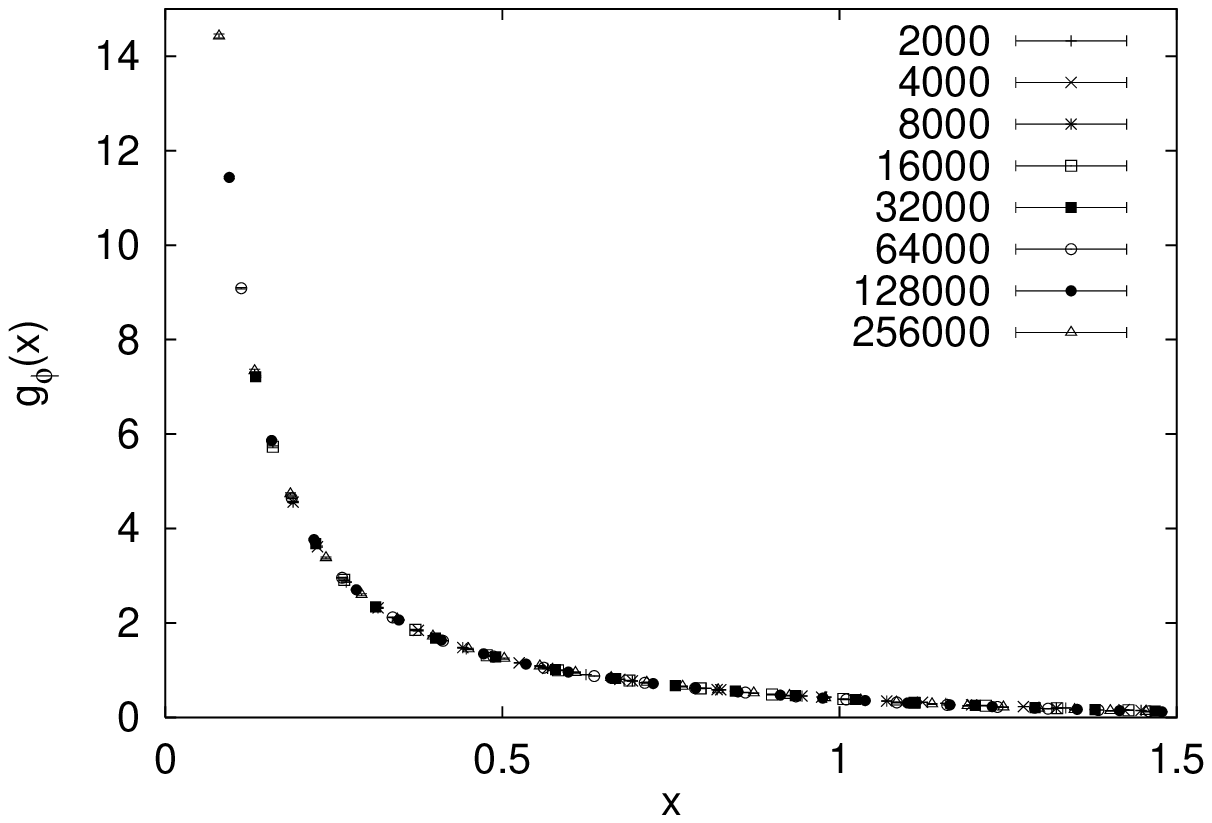}}
\centerline{   \epsfxsize=4.0in \epsfysize=2.67in 
\epsfbox{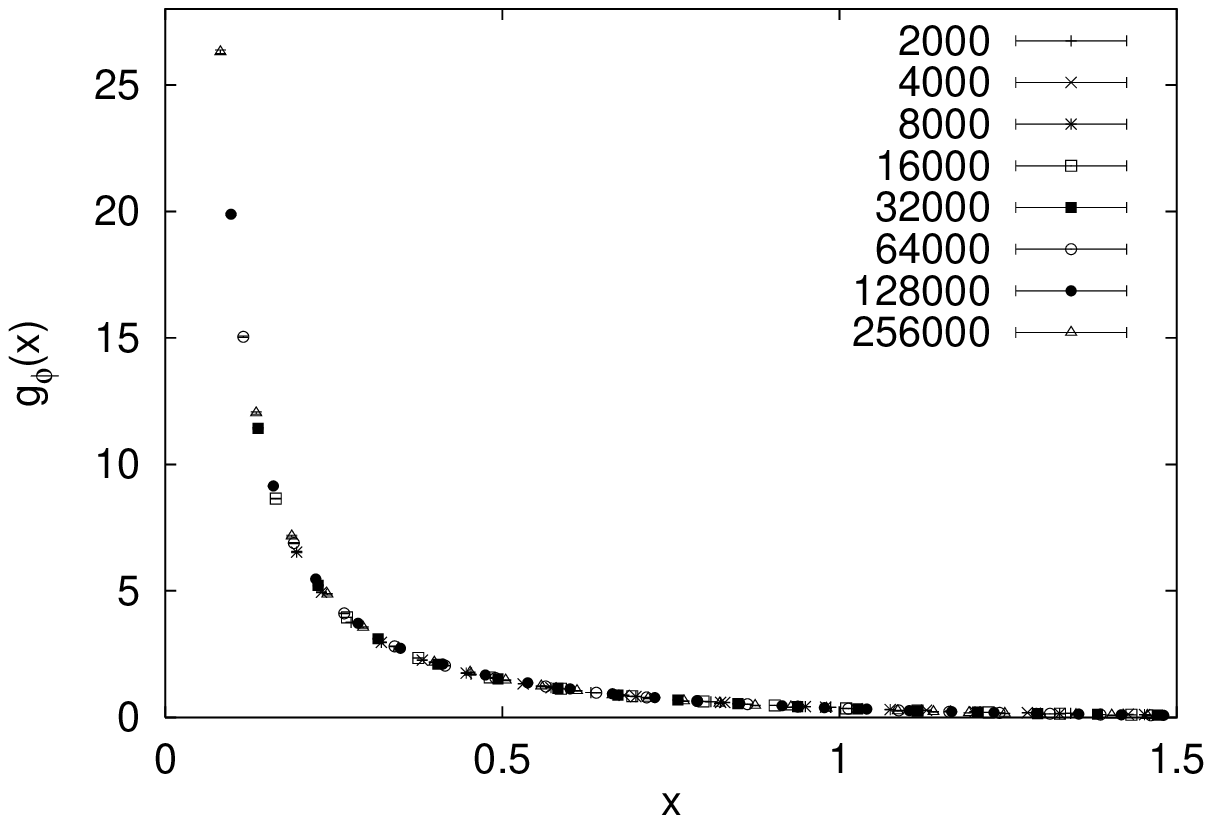}}
\caption{({\it a}) The rescaled according to Eq.~\rf{iii7}
normalized spin{--}spin correlation function 
$g_\varphi(r;N)$ for the Ising model coupled to gravity. 
We use $N_T=2000${--}$256000$, $d_h=4.0$, $a=0.51$. 
({\it b}) Same for the three{--}states Potts model coupled to
gravity. We use $a=0.55$.}
\label{f:4}
\end{figure}

\begin{figure}[htb]
\centerline{   \epsfxsize=4.0in \epsfysize=2.67in 
\epsfbox{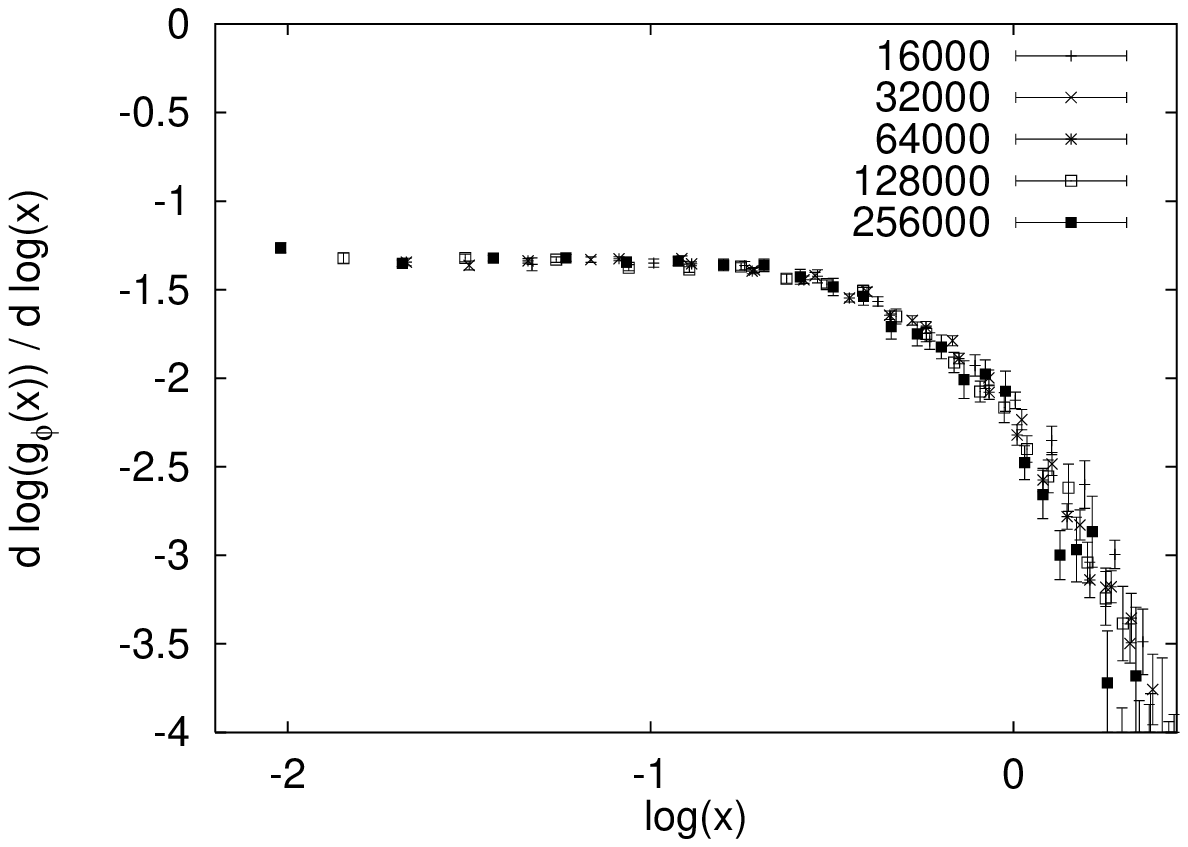}}
\centerline{   \epsfxsize=4.0in \epsfysize=2.67in 
\epsfbox{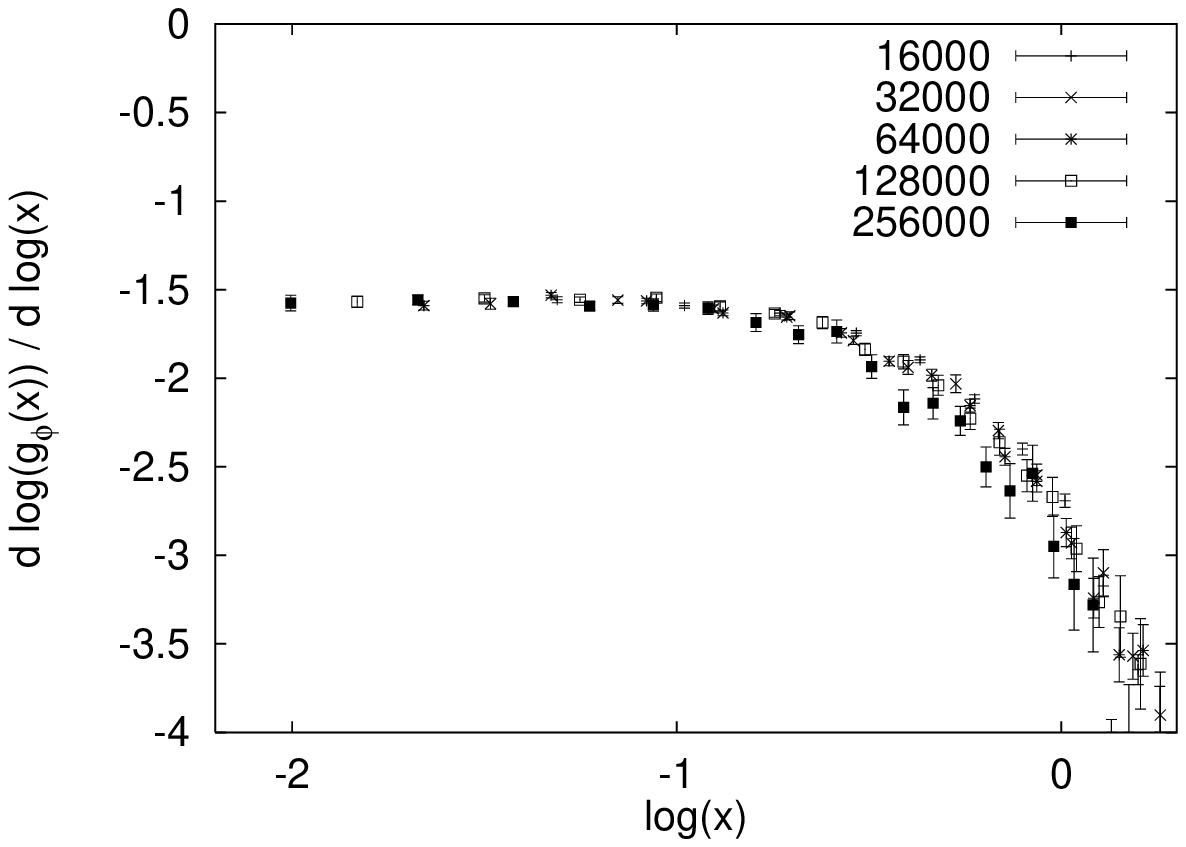}}
\caption{({\it a}) The small $x$ behaviour of the logarithmic
derivative of the rescaled normalized spin{--}spin correlation function 
$g_\varphi(r;N)$ for the Ising model coupled to gravity. We
use $N_T=16000${--}$256000$ and $x$ is obtained by using $d_h=4.0$,
$a=0.51$. 
({\it b}) Same for the three{--}states Potts model coupled to
gravity. We use now $d_h=4.0$ and $a=0.55$.}
\label{f:5}
\end{figure}

\begin{figure}[htb]
\centerline{   \epsfxsize=4.0in \epsfysize=2.67in 
\epsfbox{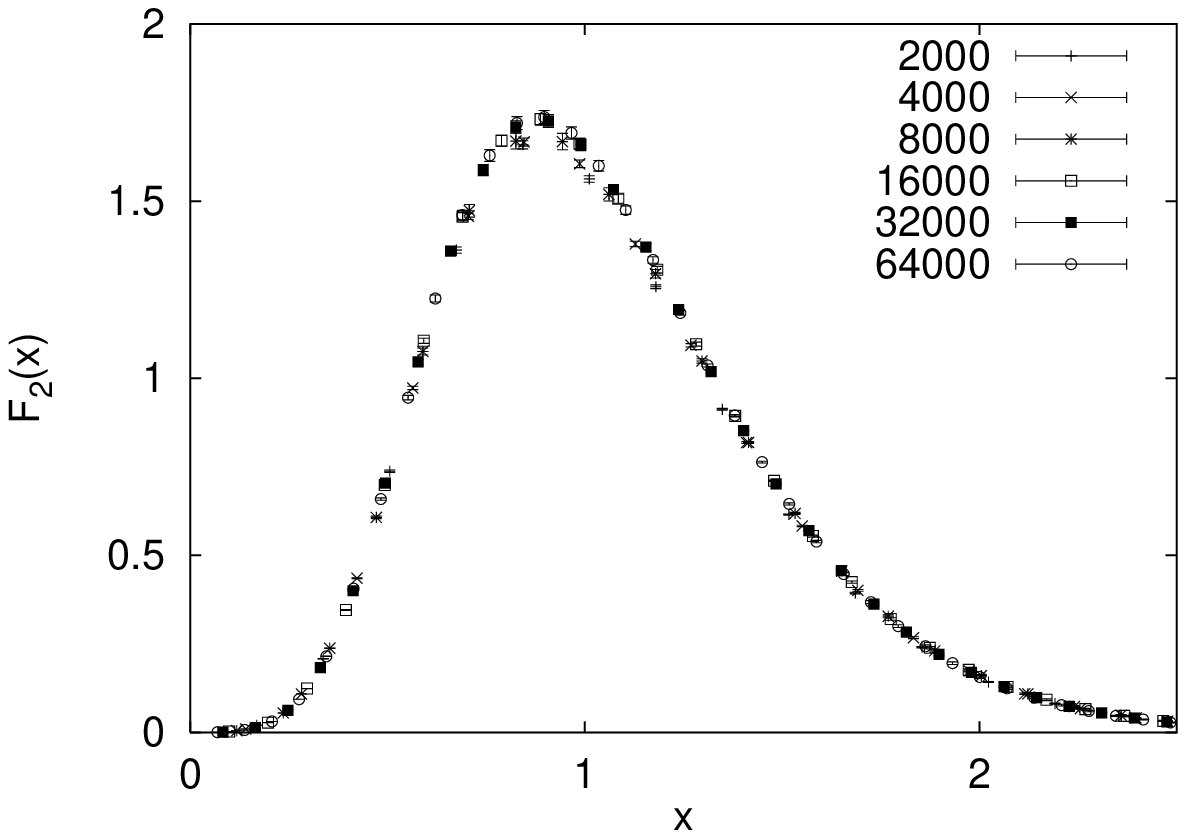}}
\centerline{   \epsfxsize=4.0in \epsfysize=2.67in 
\epsfbox{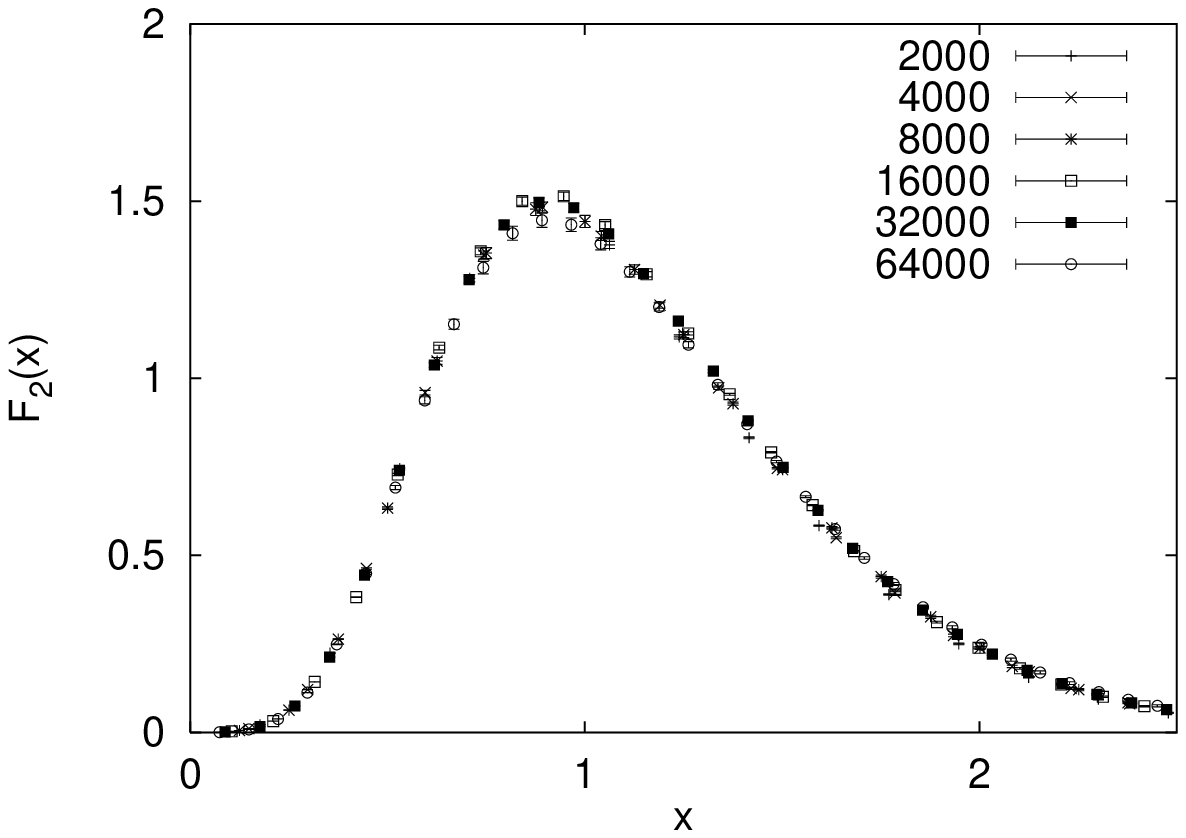}}
\centerline{   \epsfxsize=4.0in \epsfysize=2.67in 
\epsfbox{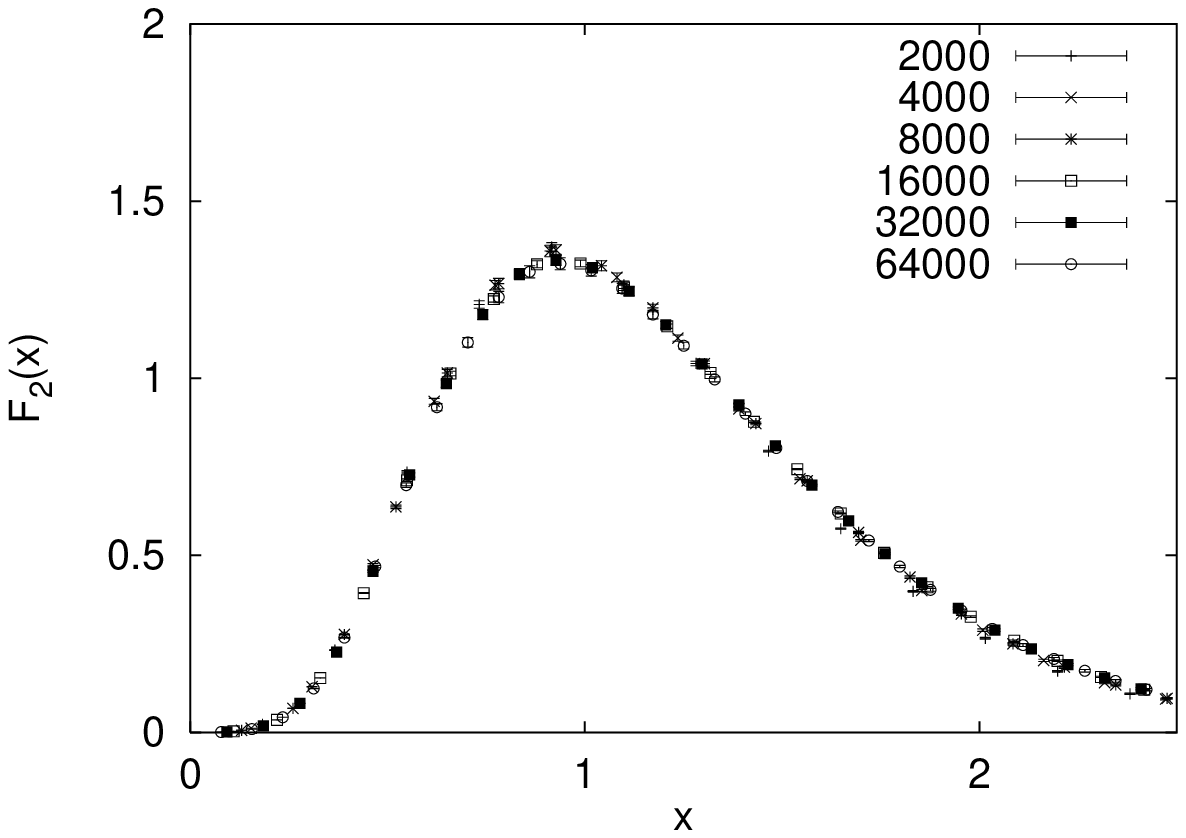}}
\caption{
({\it a}) The $\vev{l^2(r)}_N$ distributions for pure gravity rescaled
according to Eq.~\rf{iii10} using $d_h=3.88$, $a=0.00$.
({\it b}) Same as in ({\it a}) for the Ising model coupled to
gravity. $x$ is scaled using $d_h=3.99$, $a=0.0$.
({\it c}) Same as in ({\it b}) for the three{--}states Potts model
coupled to gravity. $x$ is scaled using $d_h=4.07$, $a=0.0$.}
\label{f:6}
\end{figure}

\begin{figure}[htb]
\centerline{   \epsfxsize=4.0in \epsfysize=2.67in
\epsfbox{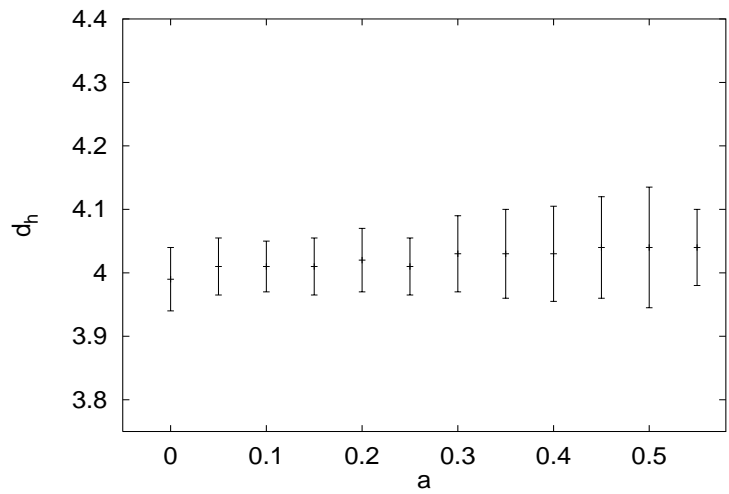}}
\caption{$d_h(a)$ from collapsing $\vev{l^2(r)}_N$ for
$N_T=16000${--}$64000$ for the Ising model coupled to gravity.}
\label{f:7}
\end{figure}

\begin{figure}[htb]
\centerline{\epsfxsize=4.0in \epsfysize=2.67in 
\epsfbox{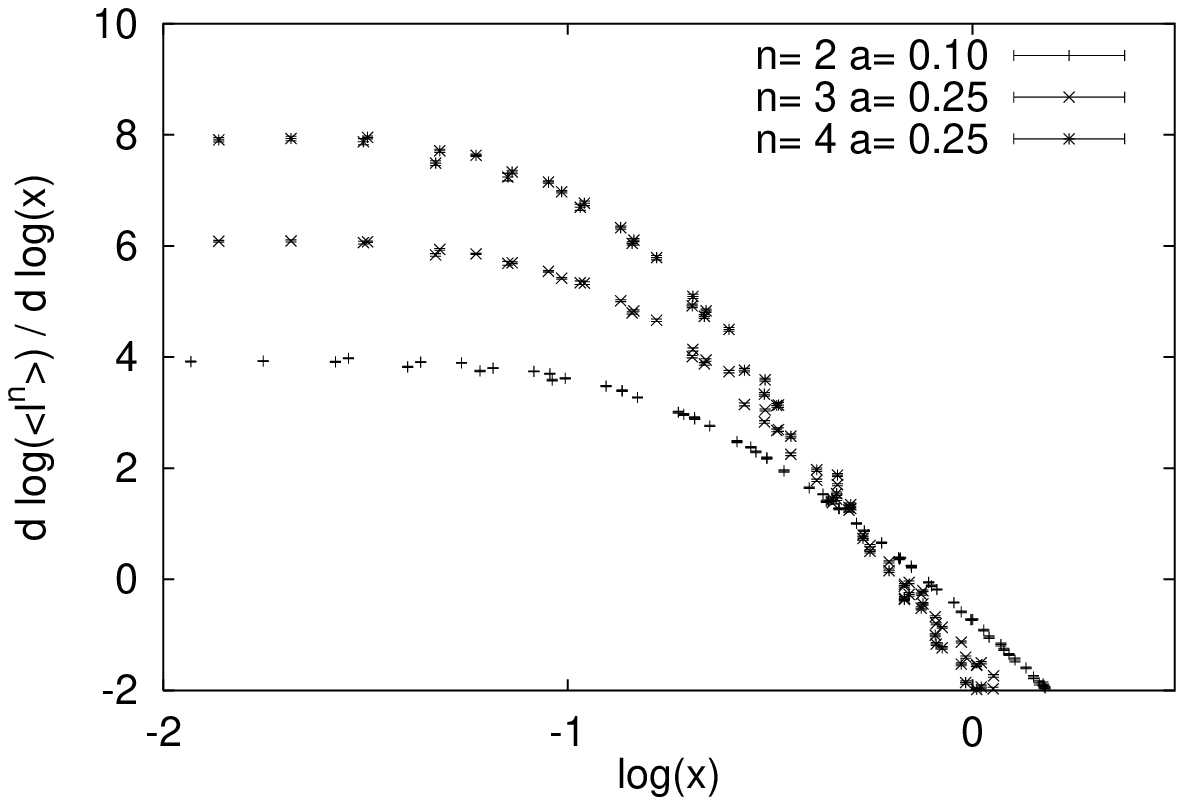}}
\centerline{   \epsfxsize=4.0in \epsfysize=2.67in 
\epsfbox{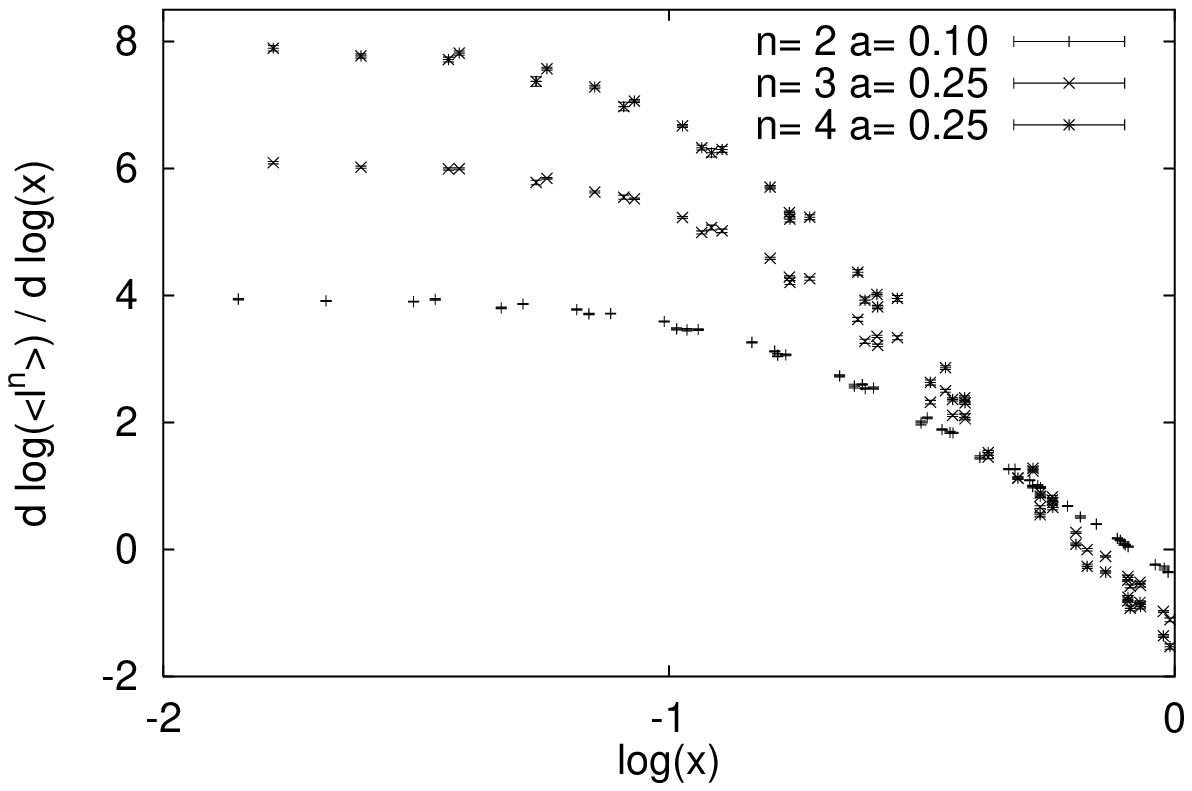}}
\centerline{   \epsfxsize=4.0in \epsfysize=2.67in 
\epsfbox{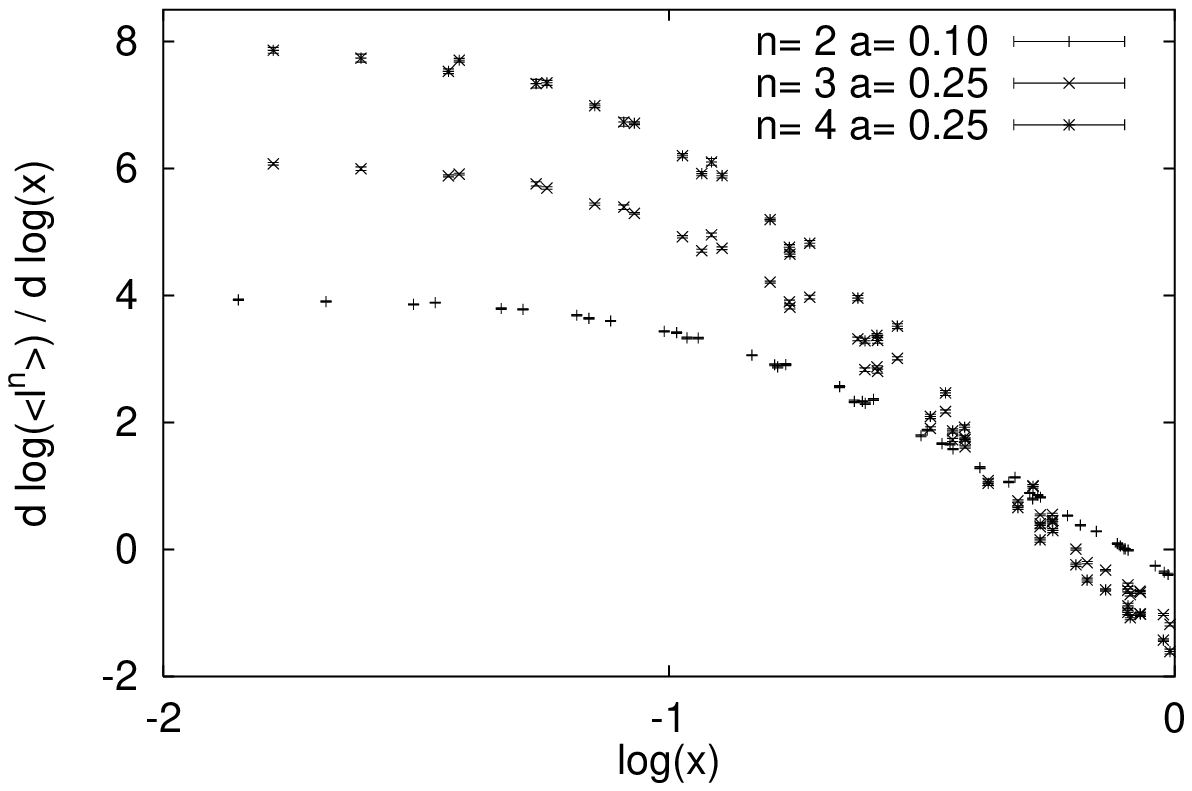}}
\caption{({\it a})The small $x$ behaviour of the logarithmic
derivative for pure gravity for $\vev{l^n(r)}_N$ for
$N_T=2000${--}$64000$.  ({\it b}) Same as in ({\it a}) for the Ising
model coupled to gravity and ({\it c}) for the three{--}states Potts
model coupled to gravity.}
\label{f:8}
\end{figure}

\begin{figure}[htb]
\centerline{\epsfxsize=4.0in \epsfysize=2.67in 
\epsfbox{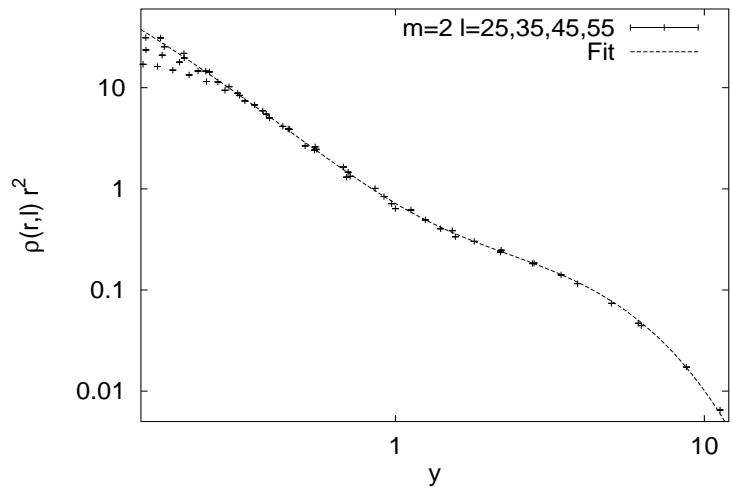}}
\caption{The loop length distribution function for pure gravity for
$N_T=64000$. The dashed line is a fit to Eq.~\rf{*in35}.}
\label{f:9}
\end{figure}

\begin{figure}[htb]
\centerline{\epsfxsize=4.0in \epsfysize=2.67in 
\epsfbox{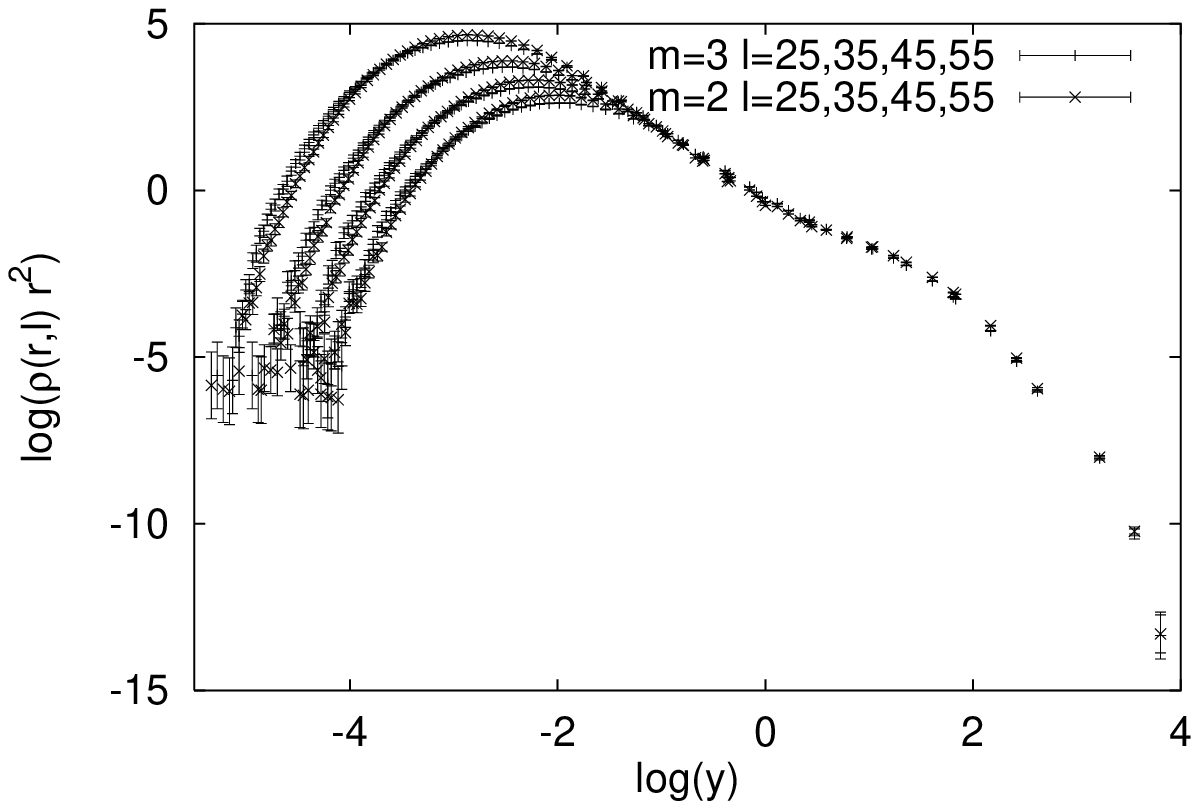}}
\centerline{\epsfxsize=4.0in \epsfysize=2.67in
\epsfbox{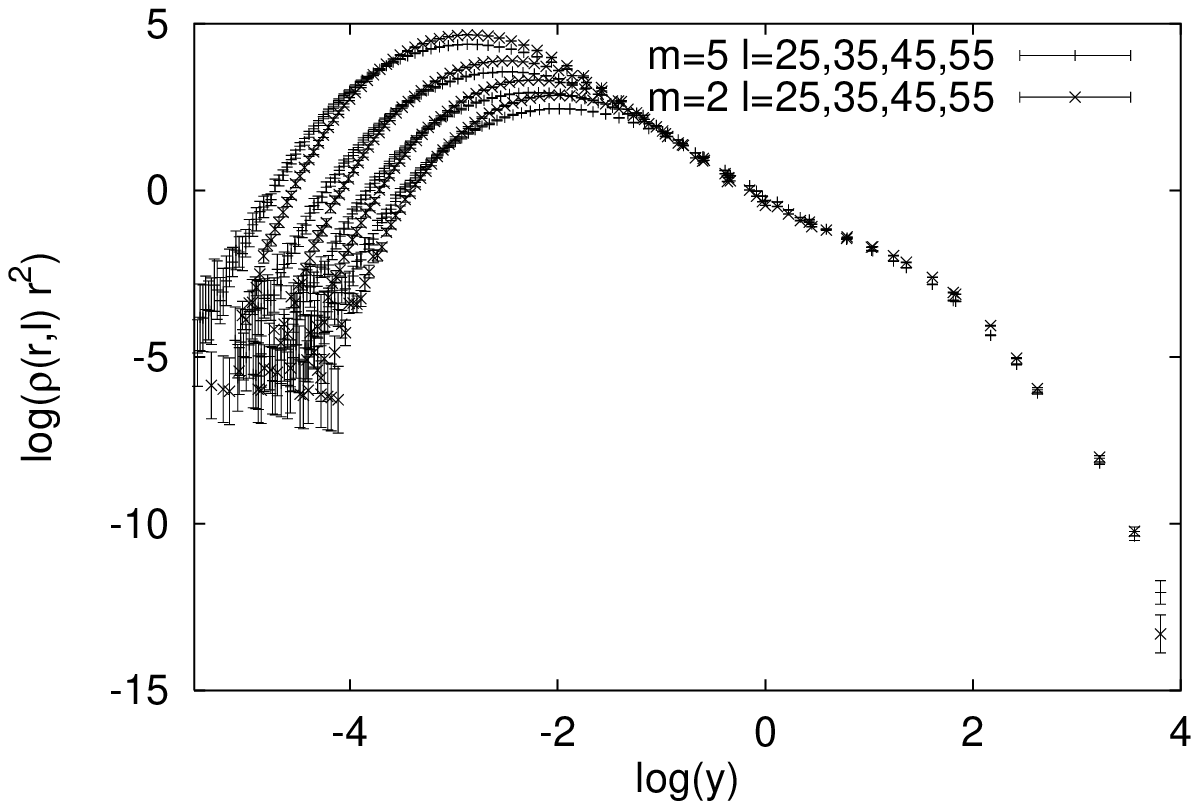}}
\caption{({\it a}) The loop length distribution function for the Ising
model for $N_T=64000$ shown together with the pure gravity one. A
wider range in $y$ is included in the plot in order to show the
$m${--}dependence in the areas where finite size effects are
important. ({\it b}) same as in ({\it a}) for the three{--}states
Potts model.}
\label{f:10}
\end{figure}

\begin{figure}[htb]
\centerline{\epsfxsize=4.0in \epsfysize=2.67in 
\epsfbox{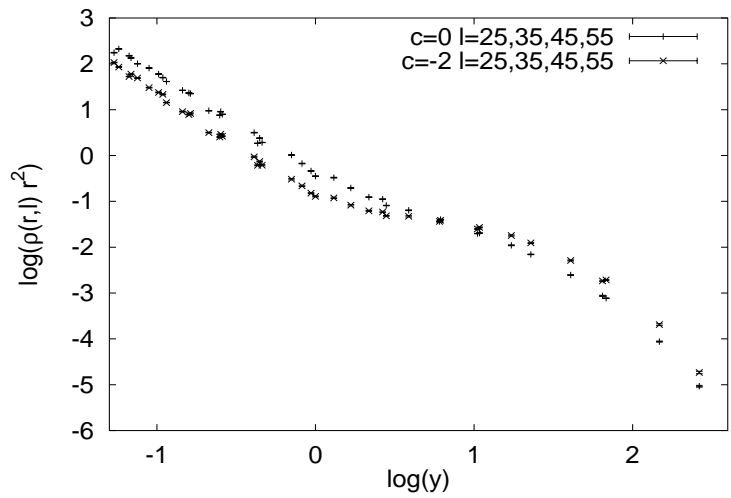}}
\caption{The loop length distribution function for pure gravity
($c=0$) and the $c=-2$ model for $N_T=64000$. The data for the $c=-2$
model is taken from \protect\cite{dgi2}.}
\label{f:11}
\end{figure}

\begin{figure}[htb]
\centerline{\epsfxsize=4.0in \epsfysize=2.67in 
\epsfbox{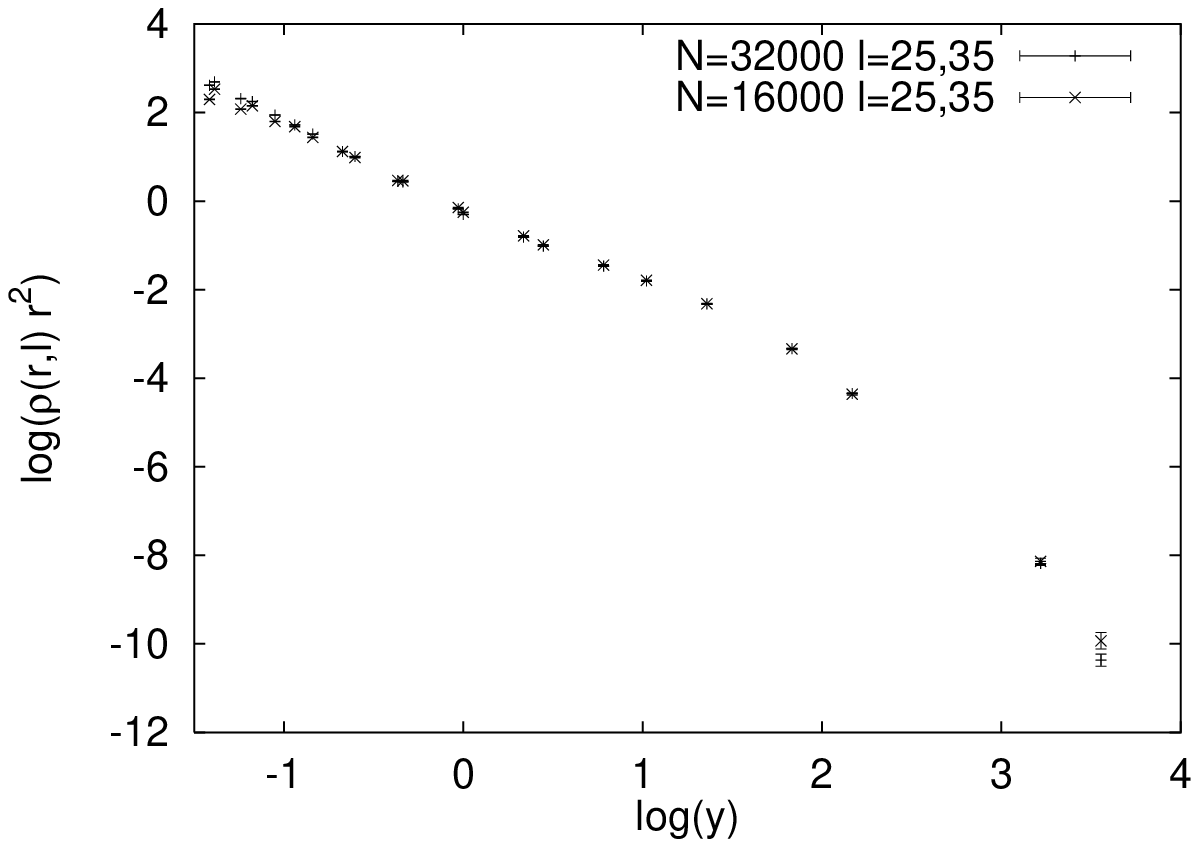}}
\centerline{\epsfxsize=4.0in \epsfysize=2.67in 
\epsfbox{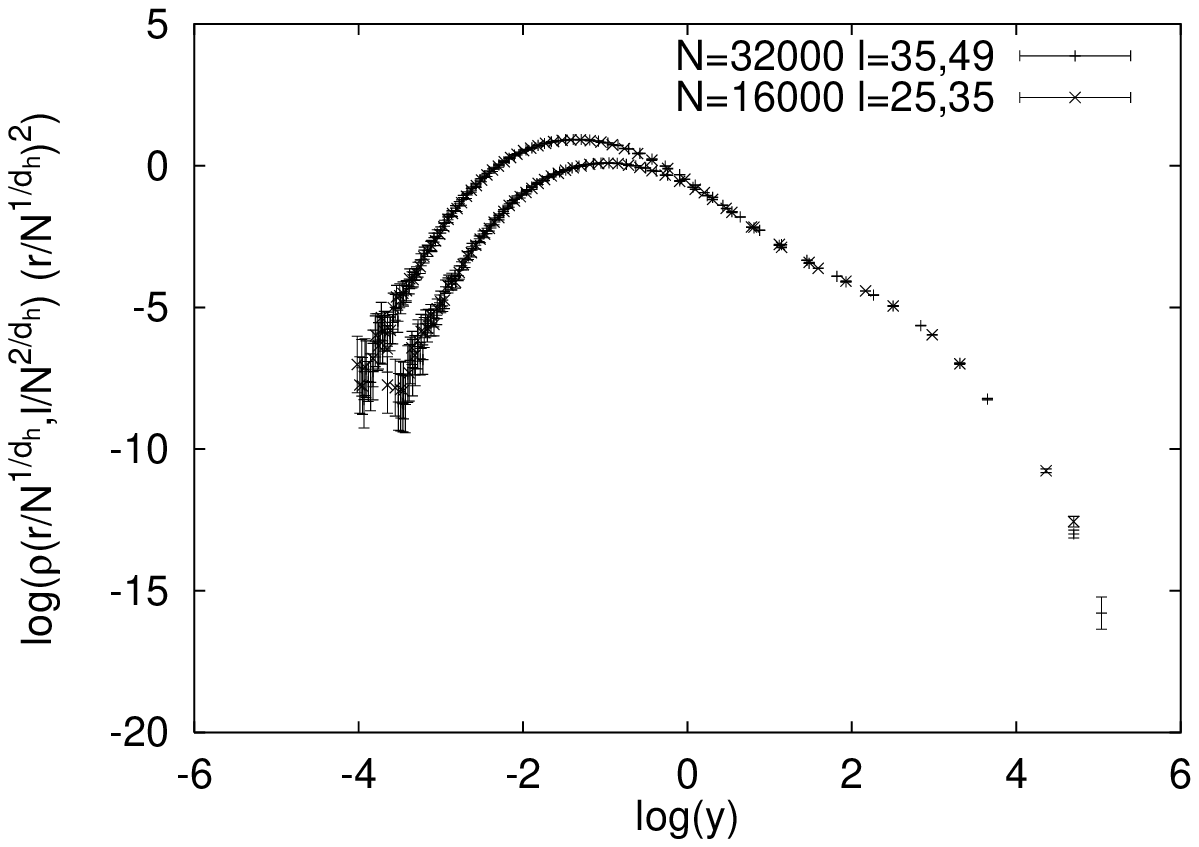}}
\caption{({\it a})The loop length distribution function for the
three{--}states Potts model coupled to gravity for $N_T=64000$ and
$32000$ for fixed $l$.  ({\it b}) The same, but for (approximately)
fixed $l/N^{2/d_h}$ for $d_h=4.0$. Similar plots are obtained for pure
gravity and the Ising model coupled to gravity.}
\label{f:12}
\end{figure}